\documentclass{aa}
\usepackage{graphicx}
\usepackage{txfonts}
\usepackage{natbib}
\usepackage{latexsym}
\usepackage{float}
\usepackage{natbib,twoopt}
\usepackage[breaklinks=true]{hyperref} %% to avoid \citeads line fills
\bibpunct{(}{)}{;}{a}{}{,}             %% natbib format for A&A and ApJ
  \newcommandtwoopt{\citeads}[3][][]{\href{http://adsabs.harvard.edu/abs/#3}%
    {\def\hyper@linkstart##1##2{}%
     \let\hyper@linkend\@empty\citealp[#1][#2]{#3}}}
  \newcommandtwoopt{\citepads}[3][][]{\href{http://adsabs.harvard.edu/abs/#3}%
    {\def\hyper@linkstart##1##2{}%
     \let\hyper@linkend\@empty\citep[#1][#2]{#3}}}
  \newcommandtwoopt{\citetads}[3][][]{\href{http://adsabs.harvard.edu/abs/#3}%
    {\def\hyper@linkstart##1##2{}%
     \let\hyper@linkend\@empty\citet[#1][#2]{#3}}}
  \newcommandtwoopt{\citeyearads}[3][][]%
    {\href{http://adsabs.harvard.edu/abs/#3}
    {\def\hyper@linkstart##1##2{}%
     \let\hyper@linkend\@empty\citeyear[#1][#2]{#3}}}
\makeatother
\newcommand\blfootnote[1]{%
 \begingroup
 \renewcommand\thefootnote{}\footnote{#1}%
 \addtocounter{footnote}{-1}%
 \endgroup}
\begin{document} 
   \title{Vortex Flow Properties in Simulations of Solar Plage Region: Evidence for their role in chromospheric heating}
   \author{N. Yadav
         \inst{1, \footnotemark}
        %  \inst{1,\footnotemark}
        %   \textbf{\large{$\ast$}}
          \and
           R. H. Cameron\inst{1}
          \and 
           S. K. Solanki\inst{1,2}
          }
   \institute{Max Planck Institute for Solar System Research,
              Justus-von-Liebig-Weg 3, 37077 G\"ottingen, Germany\\
              \email{nitnyadv@gmail.com}
         \and
             School of Space Research, Kyung Hee University, Yongin, Gyeonggi 446-701, Republic of Korea\\
             \email{solanki@mps.mpg.de}
             }
   \date{}
  \abstract
  % context heading (optional)
{Vortex-flows exist across a broad range of spatial and temporal scales in the solar atmosphere.
Small-scale vortices have been proposed to play an important role in energy transport in the solar atmosphere.
However, their physical properties remain poorly understood due to the limited spatial resolution of the observations.}
% aims heading (mandatory)
 {We aim to explore and analyze the physical properties of small-scale vortices inside magnetic flux tubes using numerical simulations, and to investigate whether they contribute to heating the chromosphere in a plage region.}
 % methods heading (mandatory)
 {Using the three-dimensional (3D) radiative magnetohydrodynamic (MHD) simulation code `MURaM', we perform numerical simulations of a unipolar solar plage region. 
  To detect and isolate vortices, we use the Swirling Strength criterion and select the locations where the fluid is rotating with an angular velocity greater than a certain threshold. 
 We concentrate on small-scale as they are the strongest and carry most of the energy.
 We explore the spatial profiles of physical quantities viz. density, horizontal velocity, etc. inside these vortices. Moreover, to apprehend their general characteristics, a  statistical investigation is performed.}
  % results heading (mandatory)
 {Magnetic flux tubes have a complex filamentary substructure harbouring an abundance of small-scale vortices.
 At the interfaces between vortices strong current sheets are formed that may dissipate and heat the solar chromosphere. 
 Statistically, vortices have higher densities and higher temperatures than the average values at the same geometrical height in the chromosphere.}
  % conclusions heading (optional), leave it empty if necessary 
 {We conclude that small-scale vortices are ubiquitous in solar plage regions, and they are denser and hotter structures that contribute to chromospheric heating, possibly by dissipation of the current sheets formed at their interfaces.}
\titlerunning{Small-scale vortices in the plage chromosphere}
\maketitle
   
\section{Introduction}
\blfootnote{ \textbf{$^\ast$} Current address: Centre for mathematical Plasma Astrophysics, Department of Mathematics, KU Leuven, Celestijnenlaan 200B, B-3001 Leuven, Belgium}
Identifying the physical mechanisms responsible for the high plasma temperatures in the solar atmosphere is a long-standing problem in solar physics.
In recent years vortices have received increasing attention due to their importance in the heating of the solar atmosphere (\citealt{Moll2011,Shelyag2011,ParkS2016, murawski2018}). 
Vortices are rotating plasma structures extending from the solar surface up to coronal heights prevalent in both, active regions and quiet Sun regions(\citealt{Wedemeyer-Bohm2012,kostas2018}).
They couple different solar atmospheric layers and are thought to excite various MHD waves, including torsional Alfv\'en waves (\citealt{Fedun2011,Shelyag2013,Jess2015,Morton2013,Morton2015}) that can heat the solar atmosphere (\citealt{Shelyag2016,2019Nature}).
They exist over a wide range of scales, with small-scale, short-lived vortices being the most abundant but not yet observed due to the limited spatial resolution of observations (\citealt{Kato2017,jiajia2018,Giagkiozis_2018,yadav2020}).

Vortices have been observed in different atmospheric layers through various photospheric and chromospheric spectral lines. Analyzing G-band observations recorded with the Swedish Solar Telescope (SST; \citealt{scharmer_2003}), \cite{Bonet2008} observed the swirling motions of bright points around intergranular points where several dark lanes converge.
They interpreted these swirls as vortex flows driven by the granulation downdrafts.
These vortices had sizes of $\sim$500 km and lifetime of the order of 5 minutes.
Large-scale ($\sim$15-20 Mm), long-lasting ($\sim$1-2 hrs) vortex flows located at supergranular junctions were identified by \cite{Attie2009} using time-series observations by the Solar Optical Telescope/Filtergraph (SOT/FG; \citealt{sot/fg}) onboard the Hinode satellite (\citealt{2007kosugi}).
Later, using observations obtained with the Imaging Magnetograph eXperiment (IMaX, \citealt{Pillet2011}) on-board the balloon-borne observatory SUNRISE (\citealt{Solanki_2010,Barthol2011,Gandorfer2011,Berkefeld2011}), \cite{steiner2010} investigated the vortex motions in the intergranular lanes and found the fine structure of numerous granules to consist of moving lanes having a leading bright rim and a trailing dark edge.
Comparing these observational results with the numerical simulations of solar surface convection, they interpreted these moving horizontal lanes as horizontal vortex tubes. 
Since the average size of the observed vortex tubes (mean radius of 150 km) was close to the spatial resolution limits of the observations and the numerical simulations, they speculated that the occurrence of such features might extend to even smaller scales.
\cite{bonet2010} used the same data set of observations but a different analysis technique.
They detected vertically-oriented vortices and attributed their origin to angular momentum conservation of the material sinking into the intergranular lanes.
\cite{iker2018} explored the connection between a persistent photospheric vortex flow and the evolution of a magnetic element using observations acquired with Hinode. 
They found that magnetic flux concentrations occur preferably at vortex sites, and the magnetic features often get weakened and fragmented after the vortex disappears. 
Their study supports the idea that a magnetic flux tube becomes stable when surrounded by a vortex flow (\citealt{schuessler1984}) and thus can support various MHD wave modes.
\begin{figure}
  \centering
  \includegraphics[scale=0.32,trim=2.5cm 0 1cm -1cm]{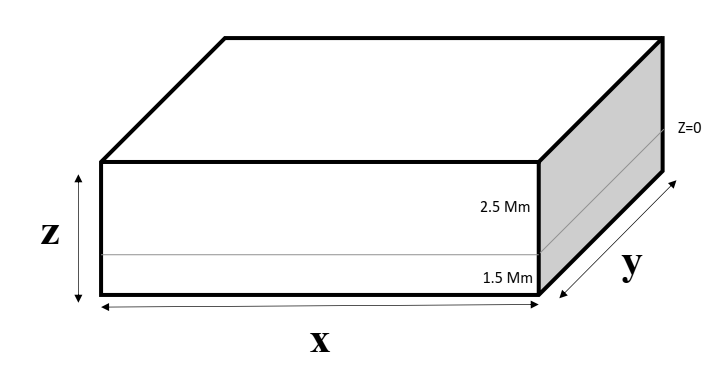}
      \caption{Sketch of the simulation box setup displaying the location of the mean solar surface at z=0 (horizontal grey line).}
         \label{Fig1}
  \end{figure}
     \begin{figure*}
  \centering
\includegraphics[scale=0.9]{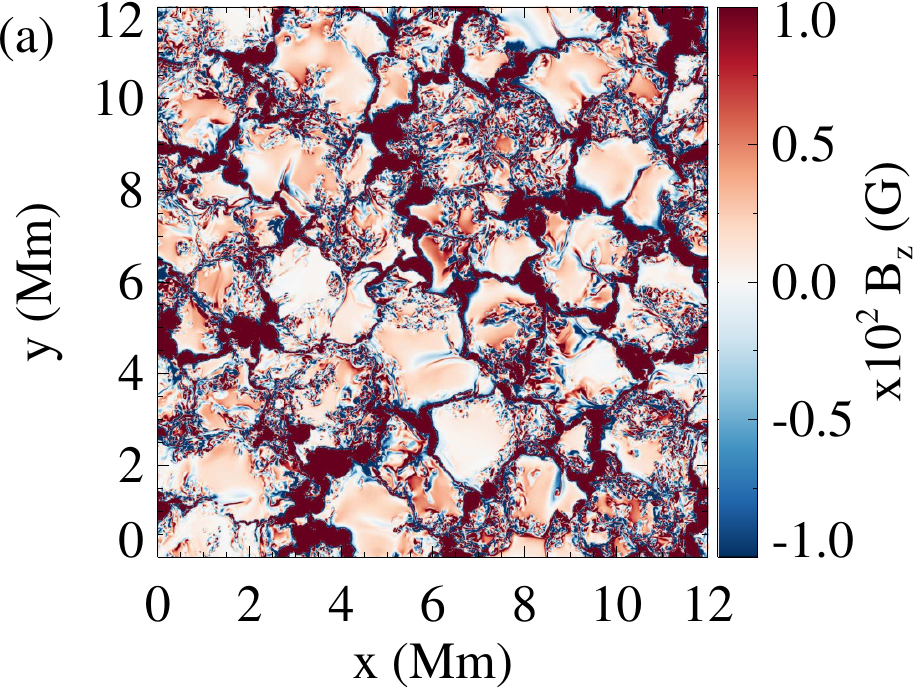}
\includegraphics[scale=0.9]{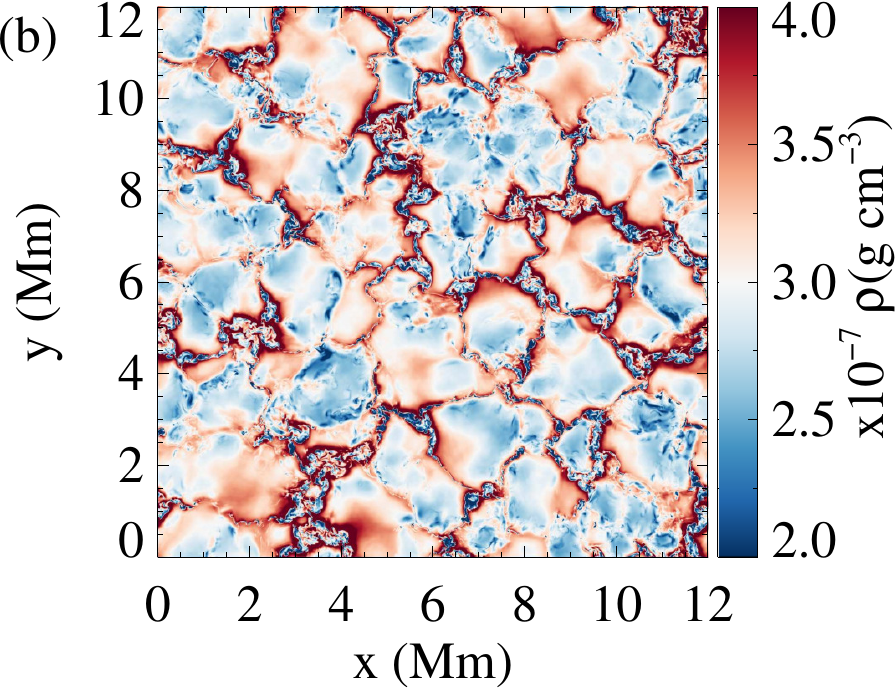}\\
\includegraphics[scale=0.9]{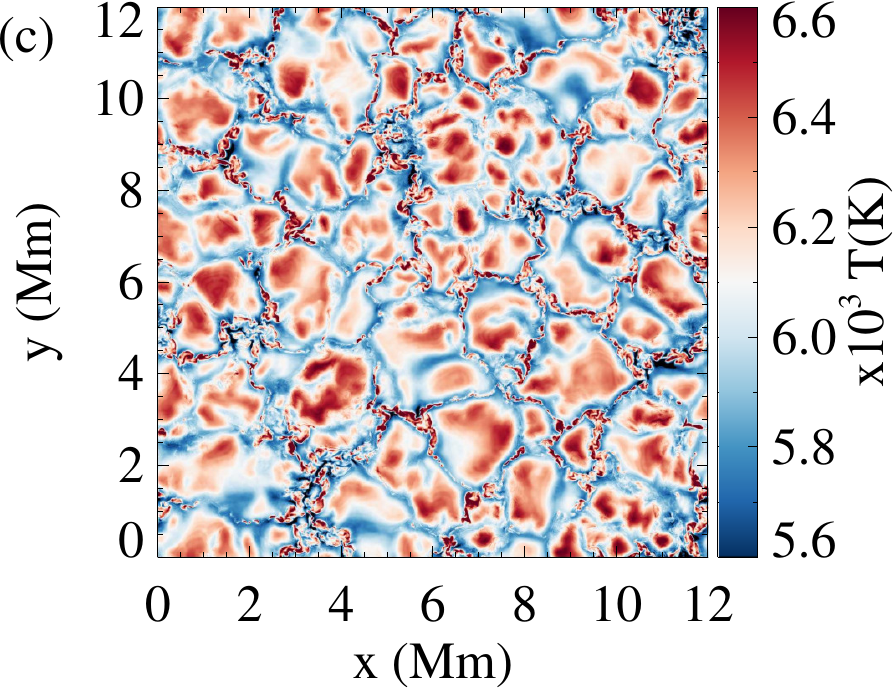}
\includegraphics[scale=0.9]{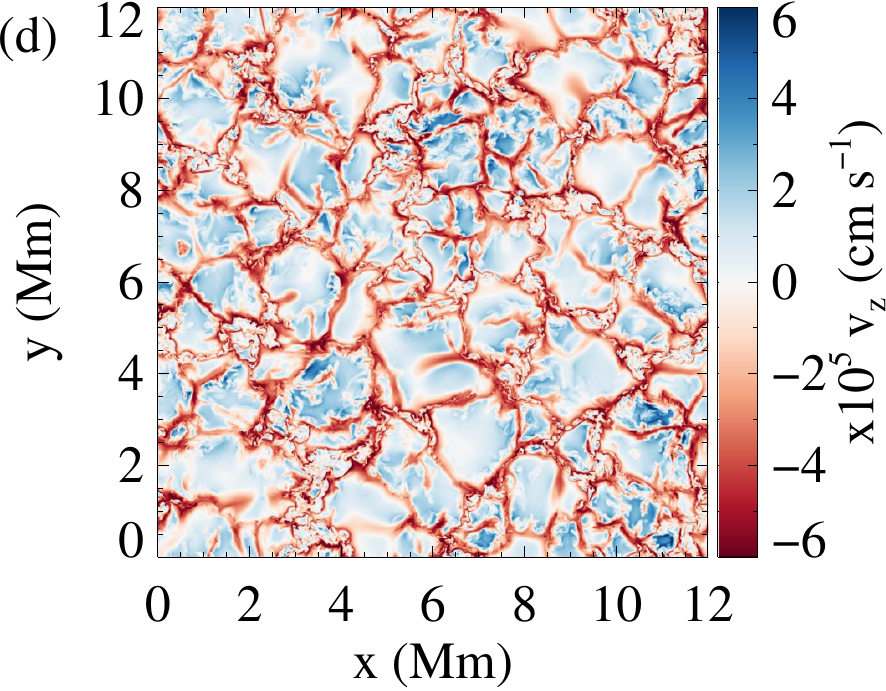}
        \caption{Spatial maps of (a): Vertical component of magnetic field ($B_z$), (b): Mass density ($\rho$), (c): Temperature (T) and (d): Vertical component of velocity ($v_z$) at the $\tau = 1$ layer, where $\tau$ is the continuum optical depth at 500 nm.}
         \label{Fig3}
  \end{figure*}
      \begin{figure*}
  \centering
  \includegraphics[scale=0.9, trim=0.cm 0.5cm 0.cm 10.5cm]{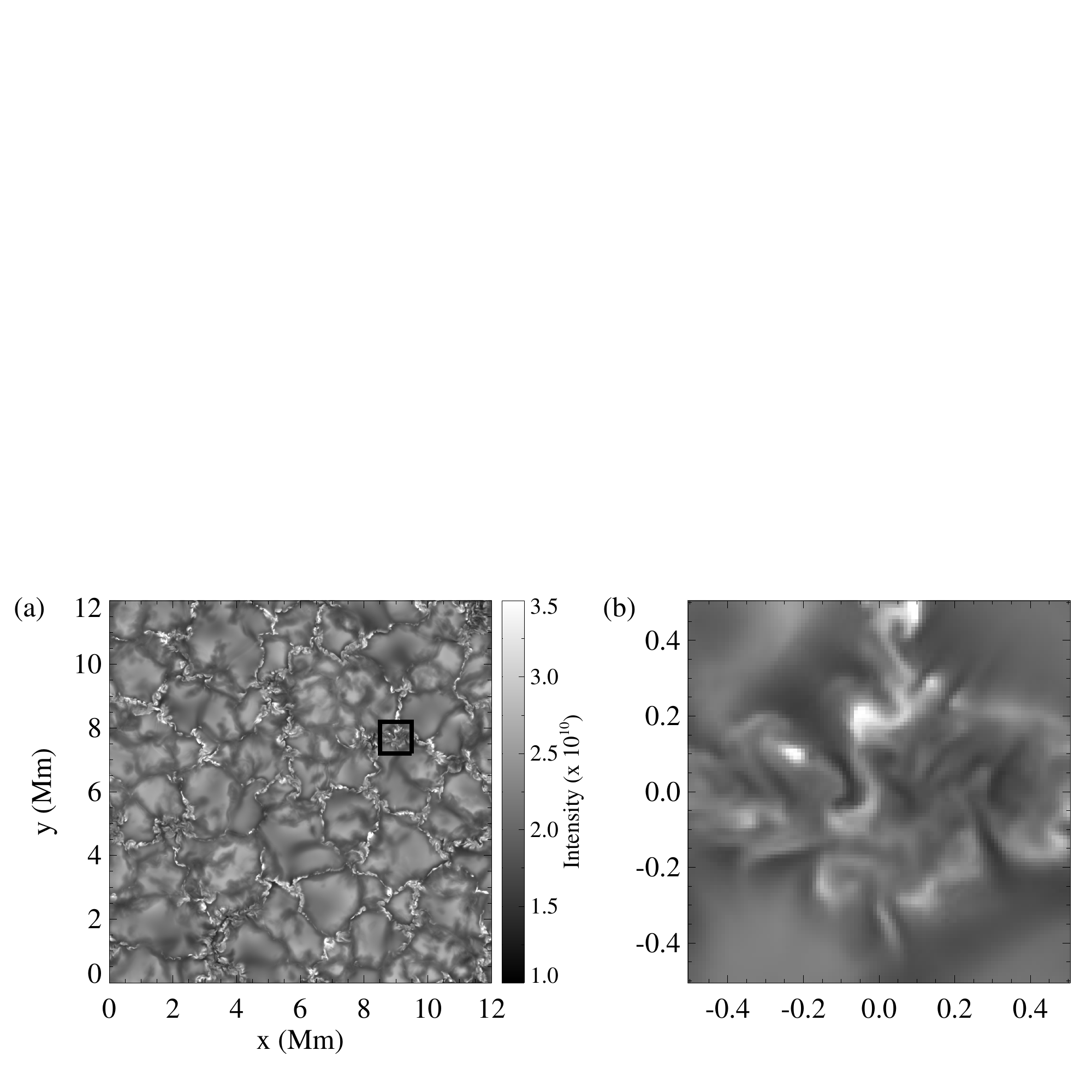}
     \caption{(a): Bolometric intensity map over the simulation domain and (b): zoomed into a magnetic element.}
         \label{Fig4}
    \end{figure*}
    \begin{figure*}
   \centering
   \includegraphics[scale=0.45,trim=8cm 0 5cm 0]{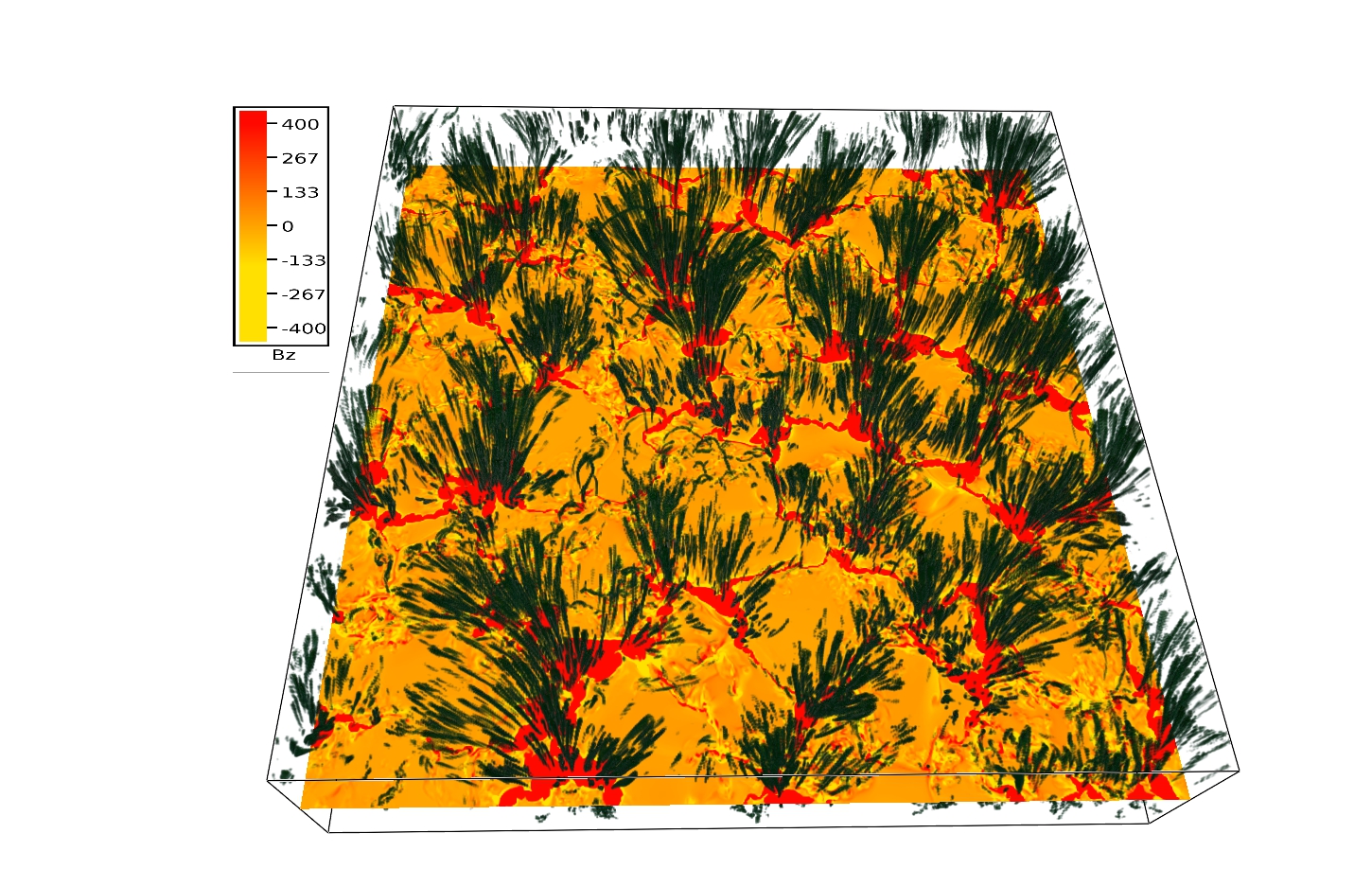}
      \caption{3D visualization of vortices in a sub-domain of the simulation box (dimension: 12 Mm x 12 Mm x 1.4 Mm, with the mean solar surface in the center). The corrugated $\tau$ =1 layer (continuum formation height) is color coded according to the vertical component of magnetic field strength (see color bar on the left).Dark green structures are vortices identified via the swirling strength criterion.}
         \label{Fig2}
   \end{figure*}

Though the initial vortex studies were focused mostly on photospheric vortices, these structures are not limited to lower atmospheric layers.
Vortex flows are also rather abundant in the chromosphere.
Using observations obtained with the CRisp Imaging Spectro-Polarimeter (CRISP) at the SST \cite{Wedemeyer-Bohm2012} analyzed chromospheric spectral lines (e.g. Ca II 854.2 nm) and detected chromospheric swirls as dark and bright rotating patches.
They estimated that there are as many as 11,000 chromospheric swirls (with a typical width of 1.5-4 Mm and lasting for 5-10 minutes) at any given time and they have associated photospheric bright points.
They interpreted these observations in terms of funnel-like magnetic field structures that have increasing cross-sectional area with height and called them `Magnetic Tornadoes'.
Besides their direct detection in the photosphere and the chromosphere, their imprints were also observed in the transition region and the lower corona through various UV and EUV channels of the Atmospheric Imaging Assembly (AIA) onboard the Solar Dynamics Observatory (SDO) space mission (\citealt{aia2012}).

Vortex flows have also been detected and analyzed in (3D radiation MHD) numerical simulations.
A detailed investigation of vortices was performed by \cite{Shelyag2011} using magnetic and non-magnetic 3D numerical simulations.
Performing radiative diagnostics of the simulation data, they revealed the correspondence between the rotation of magnetic bright points at the solar surface and small-scale swirls present in the upper photosphere ($\sim$500 km above the approximate visible solar surface level) of simulations.  
\cite{Kitiashvili_2012} investigated the thermodynamic properties of magnetized vortex tubes  in quiet-Sun simulations and discussed their importance in energy transport to the chromosphere.
They found that vortex motions concentrate the magnetic fields in near-surface layers, which in turn stabilizes the vortex tubes, reducing the influence of surrounding turbulent flows.

Vortices are also related to dynamic jet-like features observed in the chromosphere.
Analyzing SST observations, \cite{depontieu2012} reported spicules as supporting torsional Alfv\'en waves that can be excited by photospheric vortices.
Using simulations, \cite{Kitiashvili2013} demonstrated that small-scale jet-like ejections in the chromosphere can be attributed to spontaneous upflows in vortex tubes.
Moreover, there is observational evidence of the association of these rotating structures with the heating of the chromospheric plasma (\citealt{ParkS2016}). 
\cite{Attie2016} analyzed the evolution of network flares and found them to occur at the junctions of network lanes, often in association with vortex flows.
In a recent numerical study, \cite{Iijima2017} have investigated vortex flows as a driving mechanism for chromospheric jets and spicules, and demonstrated that the Lorentz force of the twisted magnetic field lines plays an important role in the production of chromospheric jets.
These studies indicate that despite originating inside the downflowing intergranular lanes, vortices can support strong plasma upflows that may result in the formation of dynamic jet-like chromospheric features.
In order to confirm a relationship between vortices and chromospheric heating, a statistical investigation is required.

The main objectives of this paper are to explore the physical properties of small-scale vortices and to investigate their role in the heating of the solar chromosphere. 

The paper is organized as follows: in Sect. \ref{2} we briefly describe the details of the numerical simulations and the methodology adopted for vortex identification.
Results are presented and discussed in Sect. \ref{3}.
The main points are summarized in Sect. \ref{4}, where also conclusions are drawn.

\section{Methodology}\label{2}
\subsection{Numerical simulation: Computational setup}
We perform 3D radiation-MHD simulations using a new version of the MURaM code (\citealt{voegler2005,Rempel2014,Rempel2017}) that can be extended to greater heights, thus reducing the effect of the upper boundary.
The simulation box covers 12Mm x 12Mm x 4Mm (with the third dimension referring to the direction perpendicular to the solar surface) that is resolved by $1200\times 1200\times 400$ grid cells providing a spatial resolution of 10 km in all three directions. 
This high spatial resolution is essential to detect the small-scale vortices and to study their fine details.
In the vertical direction, it extends from 1.5 Mm beneath the mean solar surface to 2.5 Mm above it (see Fig. \ref{Fig1}).
We use periodic boundary conditions in the horizontal direction while the bottom boundary and the top boundary are kept open and closed, respectively, allowing the free in- and outflow of plasma from the bottom boundary while conserving the total mass in the computational box.
The magnetic field is vertical at the top and bottom boundaries.

We initialize the simulations by running a purely hydrodynamic case for almost 2 hrs of solar time, so that it reaches a statistically stationary state. 
Then we introduce a uniform vertical field of 200 G magnetic strength in a hydrodynamical snapshot to simulate the conditions of a unipolar plage region, and we let the system evolve again for 1.2 hrs of solar time. 
In the course of the simulations, magnetic field lines are advected by the plasma flows in the photosphere and strong magnetic elements of 1-2 kG are formed at the solar surface.
We collect the snapshots for 7 minutes at 10 s cadence for the vortex analysis.

\subsection{Vortex identification}\label{2.2}
There is no universally-accepted definition for vortices in turbulent fluids (\citealt{jeong_hussain_1995})
Though we have an intuitive understanding of vortices as 3D objects it is difficult to decide on where each ends.
As there is no objective definition of a vortex, various vortex detection criteria are in use.

In the context of solar physics, the two most frequently used vortex detection criteria in simulations are enhanced vorticity (\citealt{Shelyag2013}; \citealt{lemmerer2017}) and enhanced swirling strength (\citealt{Moll2011}; \citealt{moll2012}).  
Enhanced swirling strength (\citealt{zhou1999}) is considered to be a better detection method, as vorticity can be high even in non-rotating shear flows.
\cite{Kato2017} compared these two vortex detection methods by applying them to the data obtained from 3D numerical simulations and found that the enhanced swirling strength method is superior to the enhanced vorticity method in all aspects.
Previously, \cite{Moll2011} successfully applied the swirling strength criterion to identify small-scale vortices in quiet-sun simulations and found vortices with a spatial size $\sim$100 km in the near-surface layers of the convection zone and in the photosphere.
These results indicate that the swirling strength criterion is a method of choice for our setup, which is similar to \citet{Moll2011}, except that our simulation box reaches much higher into the solar atmosphere.

To identify vortices by the swirling strength criterion, the velocity gradient tensor, its three eigenvalues, and corresponding three eigenvectors are calculated for all grid points using all three components of velocity at each grid point. 
The grid points with complex conjugate eigenvalues are essentially the vortex locations. 
The imaginary part of their complex conjugate eigenvalues represents the `swirling strength ($\lambda_{ci}$)', which is related to the `swirling time ($\tau_{ci}$)' by the relation $\lambda_{ci}={2\pi}/{\tau_{ci}}$. 
The swirling time signifies the rotation period of a vortex, whereas, the swirling strength gives the angular velocity or angular frequency.
The eigenvector corresponding to the real eigenvalue at these locations gives the direction vector of the vortex axis.
To extract and isolate the vortices for further investigation, we select only those contiguous features that rotate significantly rapidly (i.e. which have swirling time less than 100 s) and have more than 200 grid points in 3D space in each feature to avoid smaller transient structures.
The threshold in swirling time (or swirling strength) is chosen by visual inspection of swirling strength maps at various heights such that the selected regions encompass the strongly rotating vortices.

\section{Results}\label{3}
    \begin{figure*}
    \centering
         \includegraphics[scale=0.14]{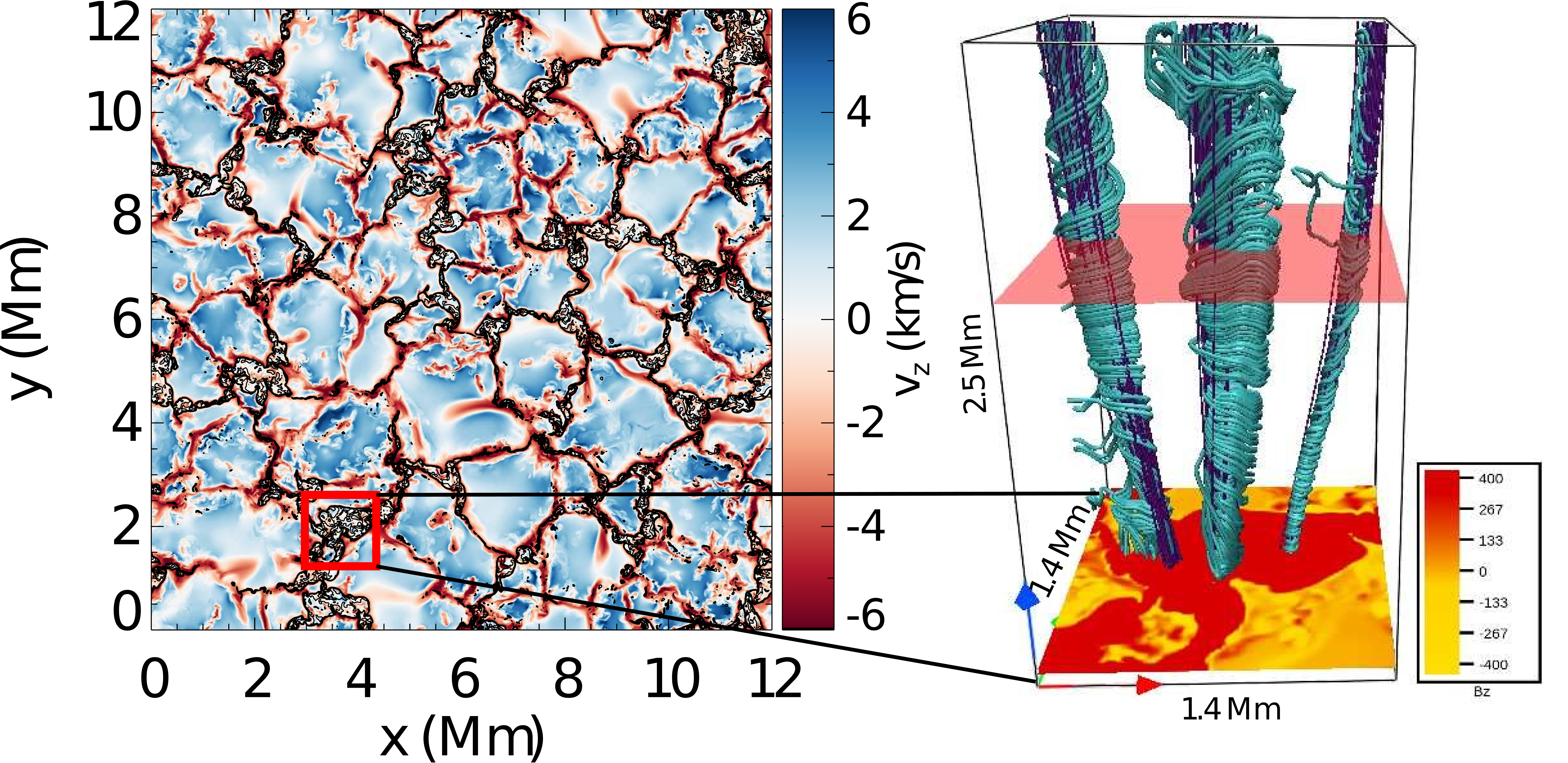}
      \caption{Left:Vertical component of flow velocity ($\mathrm{v_z}$) at the $\tau =1$ layer with black contours of $\mathrm{B_z}$ ranging from 400G to 2500G and a red box highlighting one magnetic element, which is plotted on an enlarged scale in the right panel, where the image at the bottom now shows the vertical component of the field, $B_z$, instead of $v_z$ in the left panel. Right: 3D view of a sub-domain covering the volume $1.4 \: \mathrm{Mm}\times 1.4 \: \mathrm{Mm}\times 2.5 \: \mathrm{Mm}$ corresponding to the red square. Velocity streamlines and magnetic field lines are shown in sea-green and deep purple colors, respectively, at the selected vortex locations. A perpendicular cut at 1.5 Mm above the surface is indicated by the light red plane at which vortex properties will be shown and discussed later in the paper.
    }
 \label{Fig8}
   \end{figure*}
   
Figure \ref{Fig3} displays the maps of various physical quantities at the corrugated $\tau = 1$ layer (continuum formation height).
Panel (a) displays the vertical component of the magnetic field strength with values saturated at $\pm$ 100 G. 
Small-scale mixed-polarity magnetic fields surrounding the strong magnetic field concentrations are seen.
Panel (b) shows the mass density ($\rho$), the mass density is higher in the intergranular lanes due to the build-up of excess pressure there to stop the horizontal convective flows.
However, the strong magnetic field concentrations have low mass densities due to horizontal pressure balance.
The temperature is displayed in panel (c).
Clearly, in the intergranular lanes it is highly structured.
The vertical component of velocity ($v_z$) is displayed in panel (d). 
Convection is quenched by strong magnetic fields in magnetic concentrations, hence we have almost negligible flows inside the strong magnetic elements.

The map of bolometric intensity (i.e. intensity integrated over all wavelengths) is displayed in panel (a) of Fig. \ref{Fig4}.
A small magnetic region has been enlarged and displayed in panel (b). 
Ribbon or flower-like intensity structures with bright edges can be seen in the intergranular lanes, caused by the complex interaction between convection, magnetic fields, and radiation.
Such intensity profiles have been observed previously in a unipolar ephemeral region (\citealt{narayan}).

Figure \ref{Fig2} shows a sub-domain of the simulation box (dimension: $12\; \mathrm{Mm}\times12\;\mathrm{Mm}\times1.4\;\mathrm{Mm}$, with the mean solar surface in the center of the z-axis range).
Only a restricted height range is displayed in this figure to clearly represent the foot-points of vortex features, which become volume-filling in the upper layers.
The vertical component of magnetic field strength at the $\tau$ =1 layer is displayed by a color map (the color bar is saturated at $\pm$ 400 Gauss for better visualization).
The magnetic field strength in small magnetic elements typically lie in the range $\sim$1-2 kG at the solar surface and vortices originate inside these magnetic concentrations.
Vortices are selected using the enhanced swirling strength criteria as described in Sect. \ref{2.2} and are shown in dark green.
Here, we take the swirling period threshold to be 100 s which corresponds to a swirling strength threshold of 0.0628 rad/s. 
Consequently, only those vortices are selected which rotate with an angular velocity higher than this swirling strength threshold.

\subsection{Individual vortices}\label{individual}
\begin{figure}
\centering
\includegraphics[scale=0.05,trim=0 0 0 0]{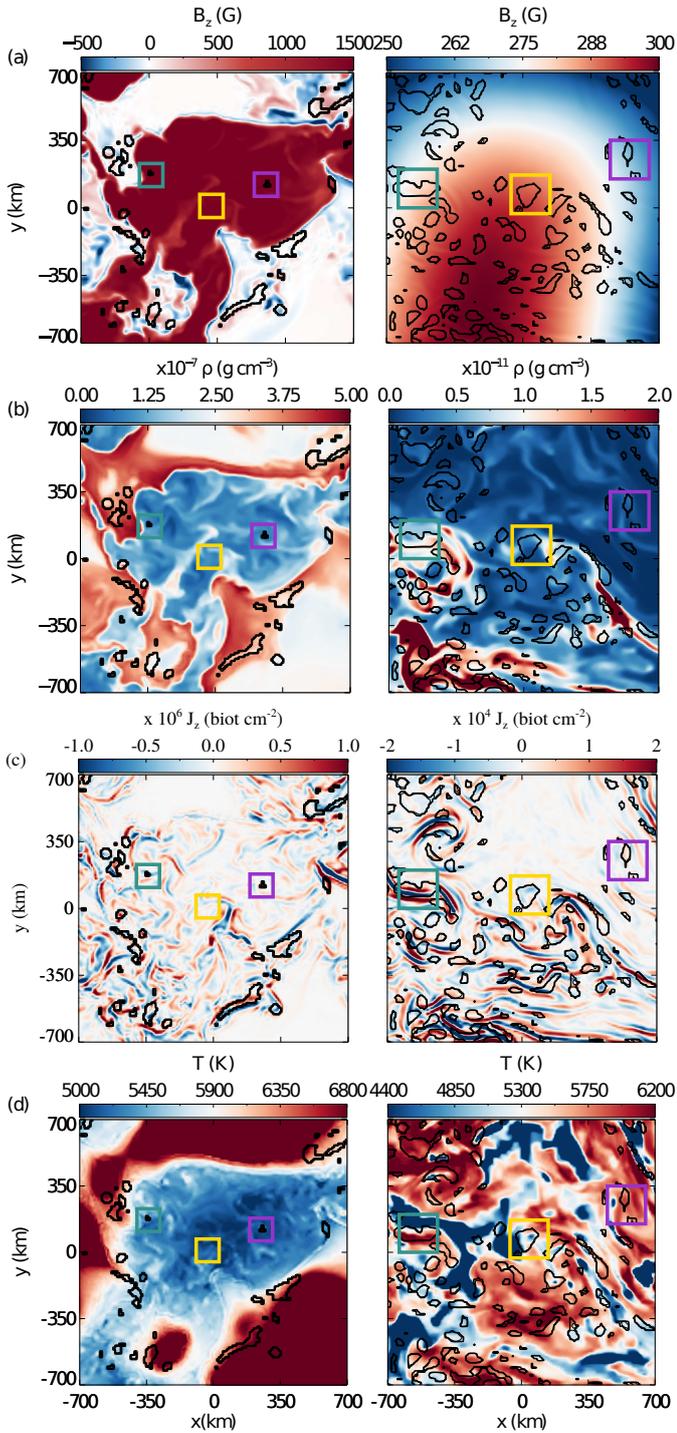}
 \caption{Maps of (a): Vertical component of the magnetic vector ($\mathrm{B_z}$), (b): Mass density ($\rho$), (c): Vertical component of current ($\mathrm{J_z}$) and (d): Temperature (T) at the mean surface (left column) and in the chromosphere (right column). Three colored squares mark the locations of the selected vortices shown in Fig. \ref{Fig8} }
   \label{Fig10}
  \end{figure}

\begin{figure}
\centering
 \includegraphics[scale=0.32,trim=2cm 3cm 4cm 0]{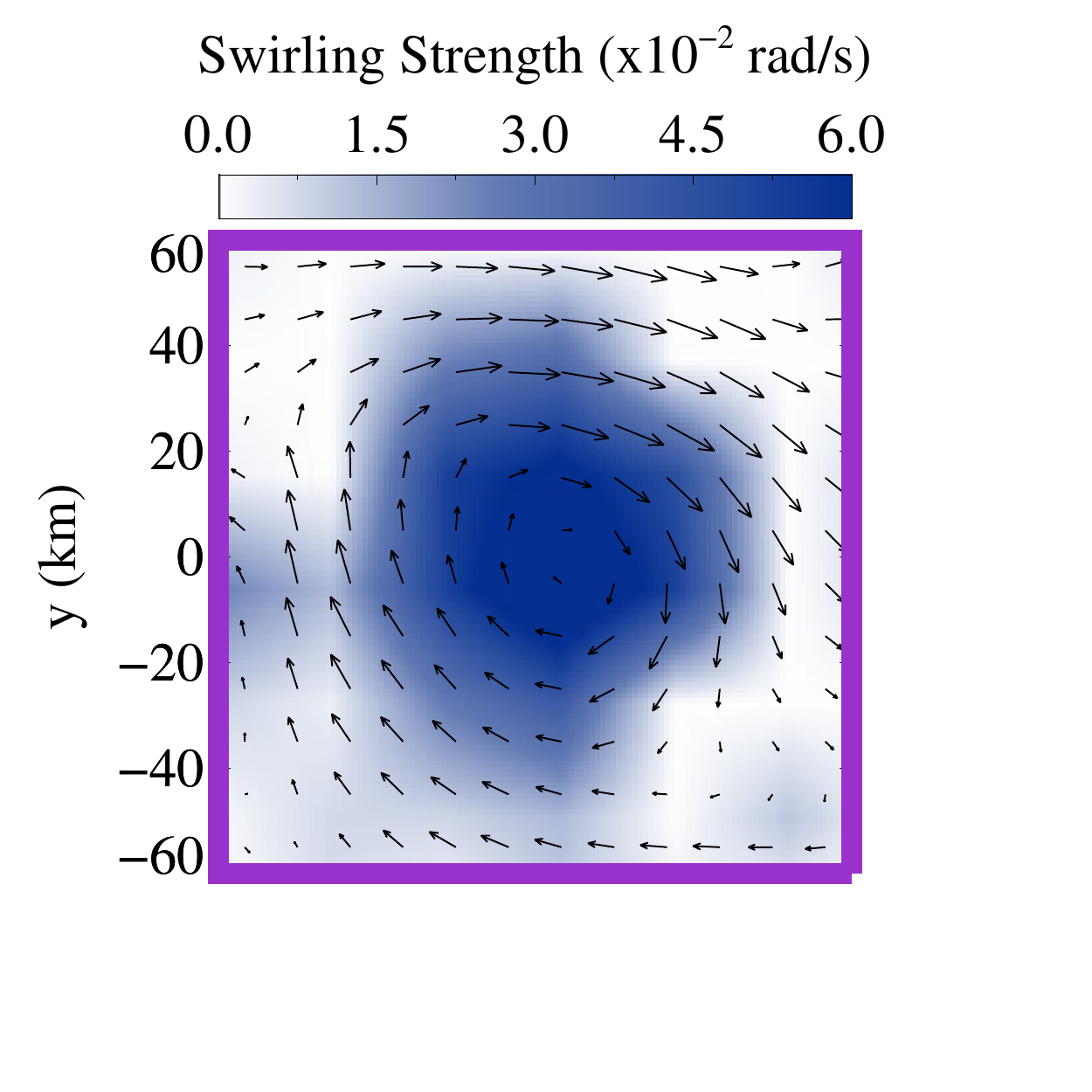}
\includegraphics[scale=0.32,trim=0 3cm 4cm 0]{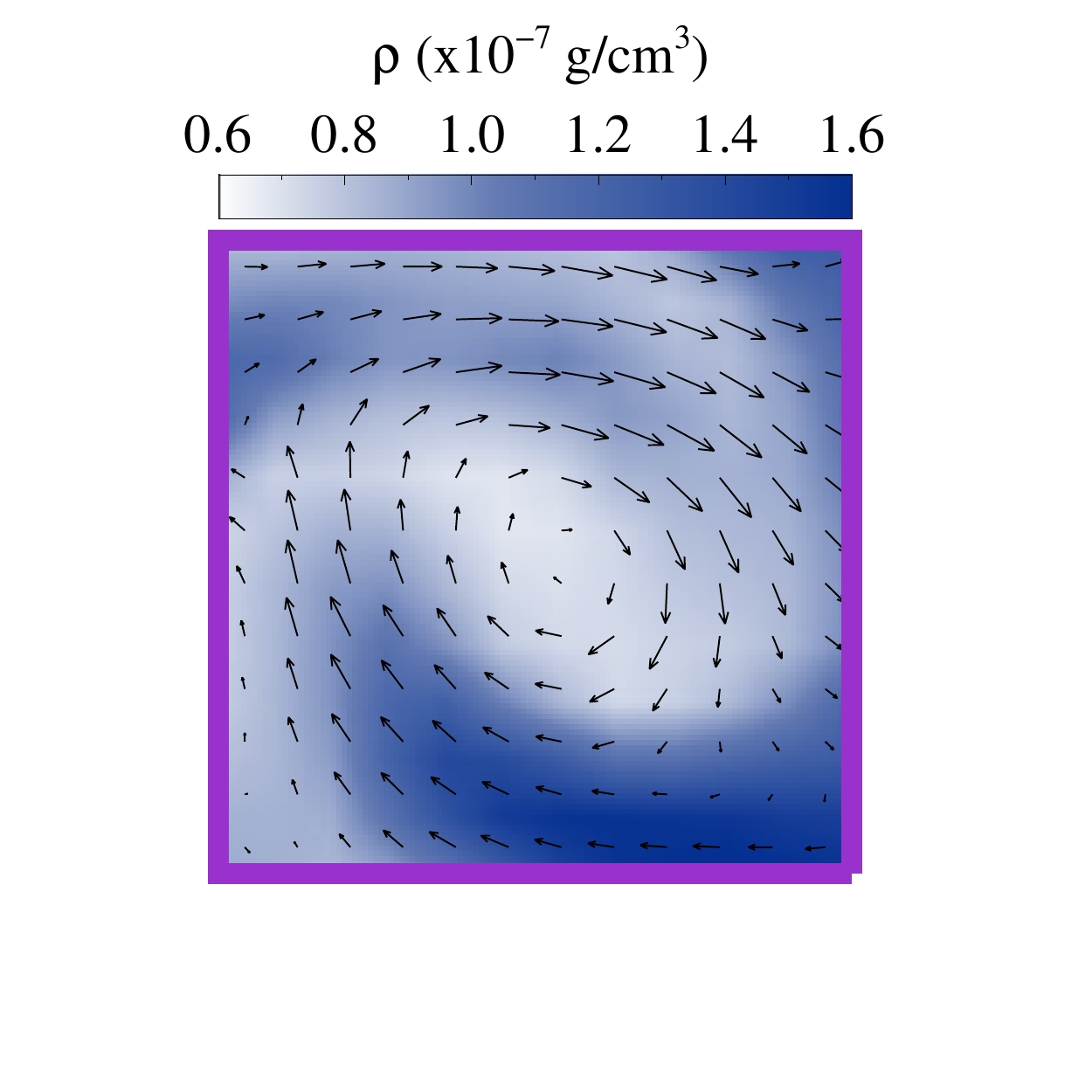}
 \includegraphics[scale=0.32,trim=0 3cm 4cm 0]{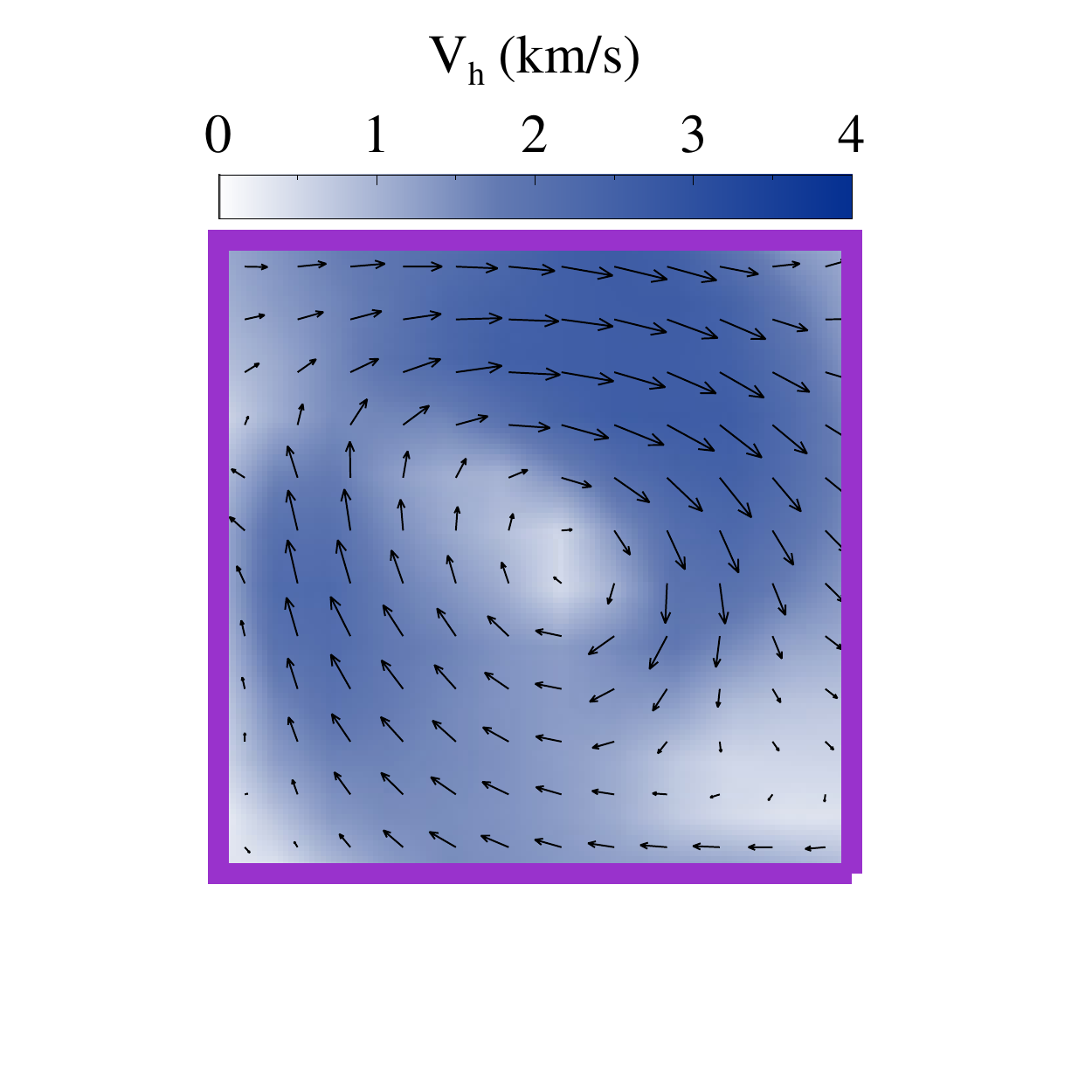}\\
\includegraphics[scale=0.32,trim=2cm 3cm 4cm 1.8cm]{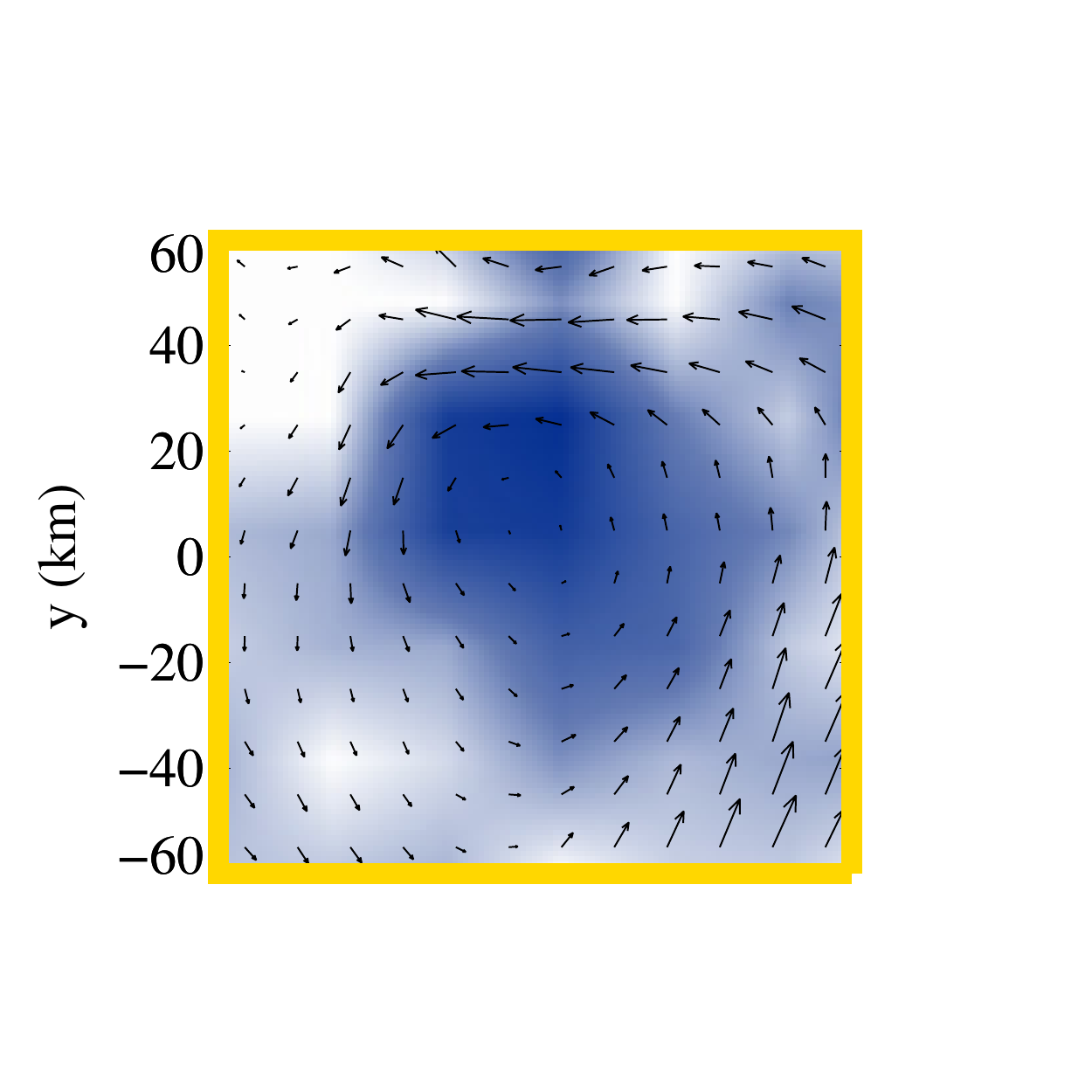}
\includegraphics[scale=0.32,trim=0 3cm 4cm 1.8cm ]{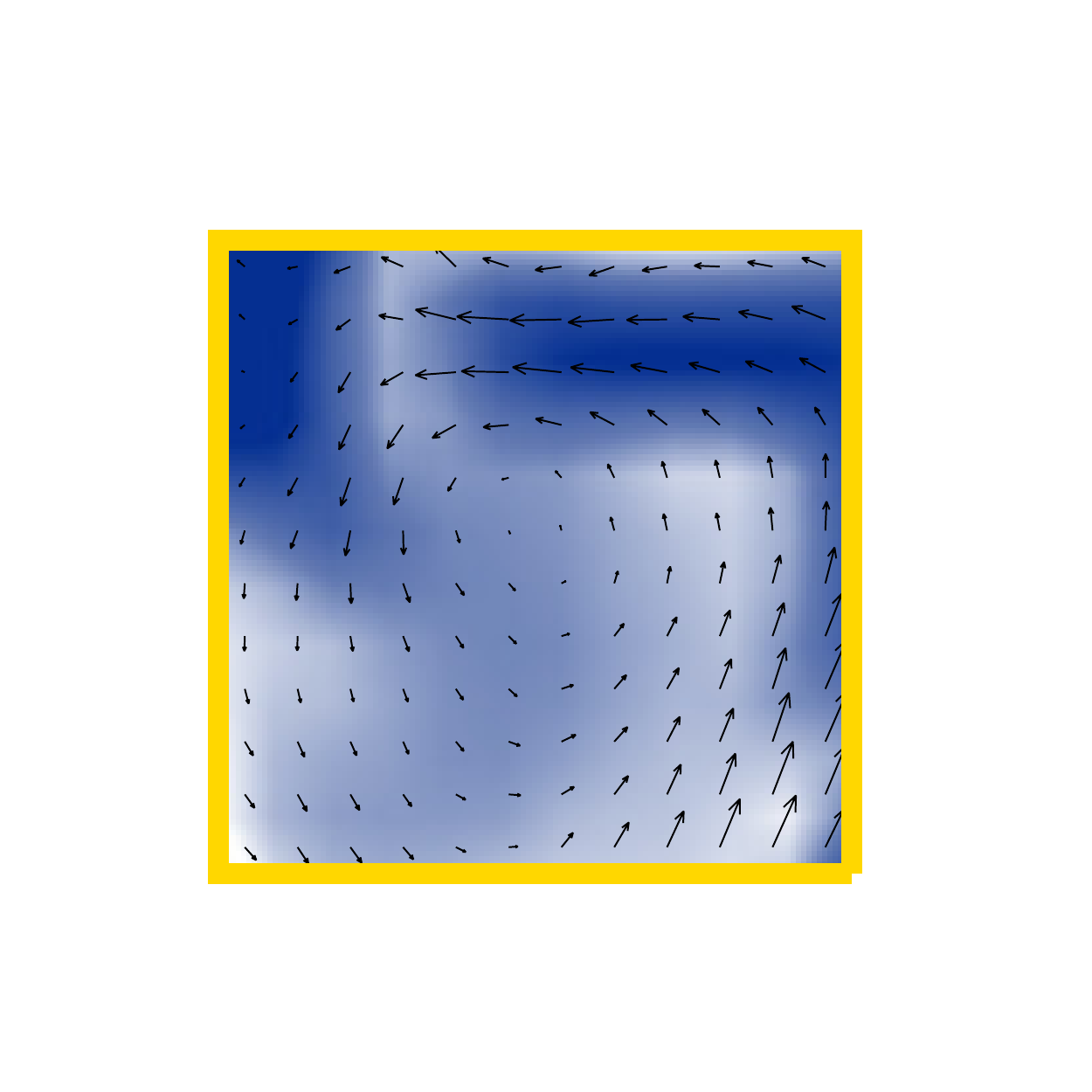}
\includegraphics[scale=0.32,trim=0 3cm 4cm 1.8cm]{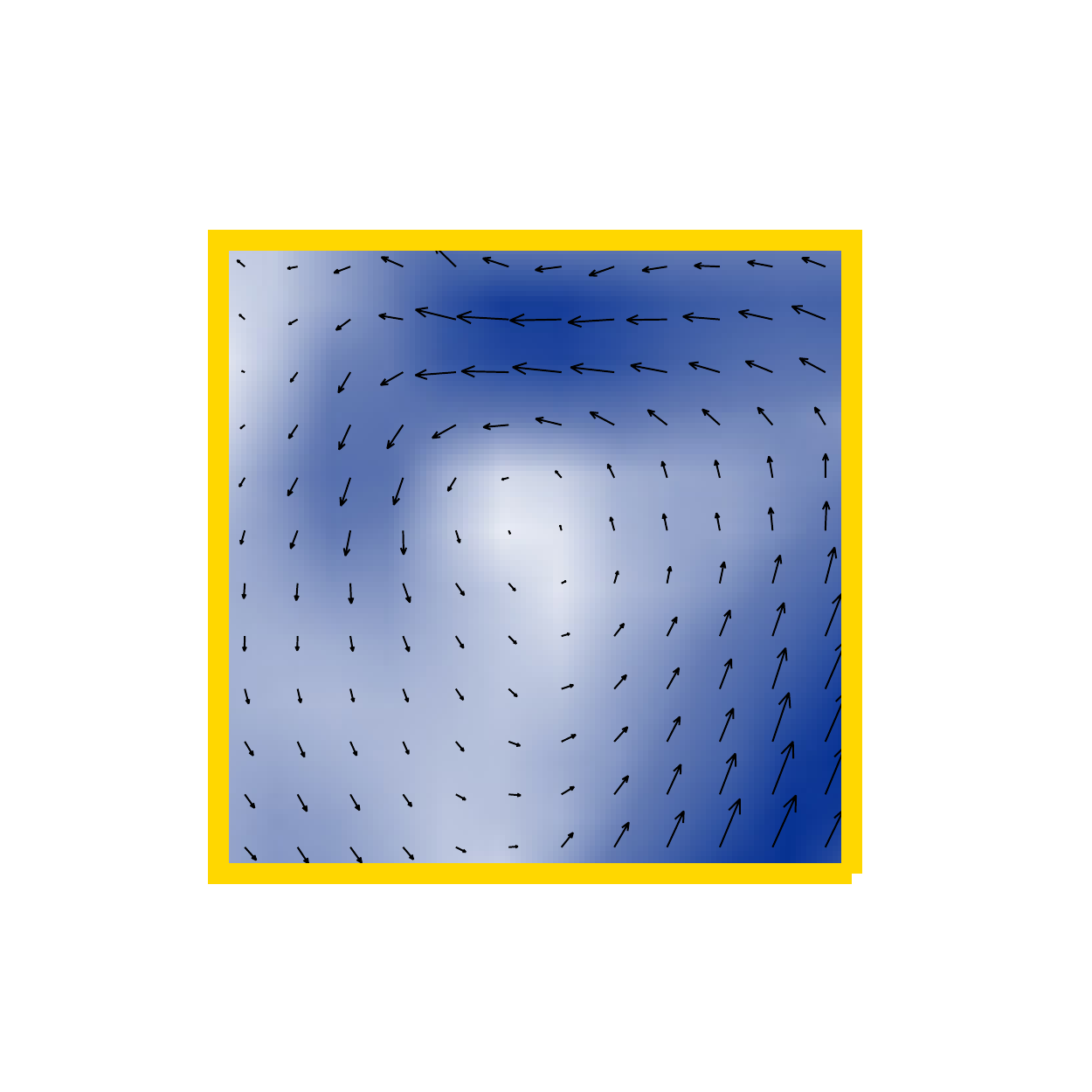}\\
\includegraphics[scale=0.32,trim=2cm 1cm 4cm 1.8cm]{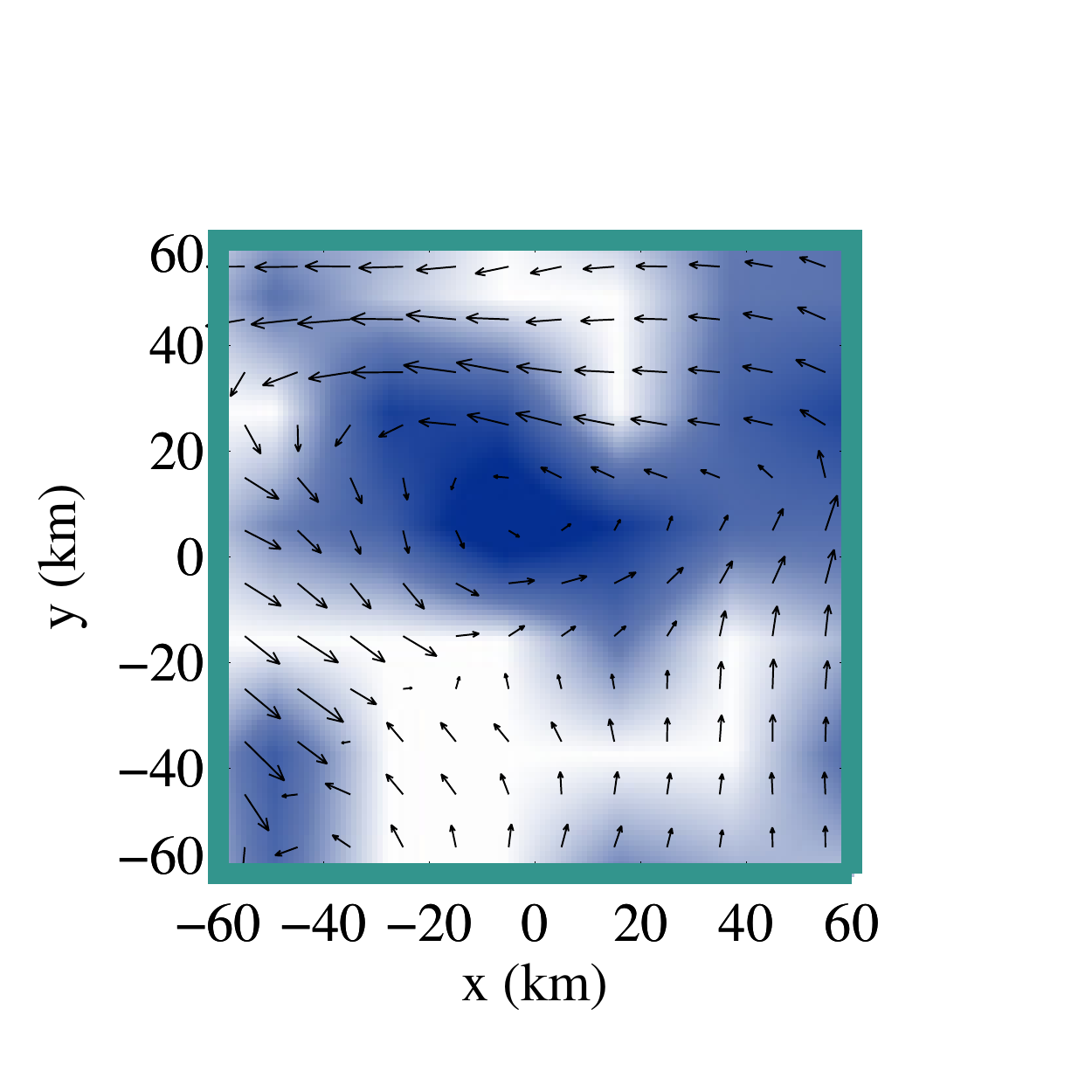}
\includegraphics[scale=0.32,trim=0 1cm 4cm 1.8cm]{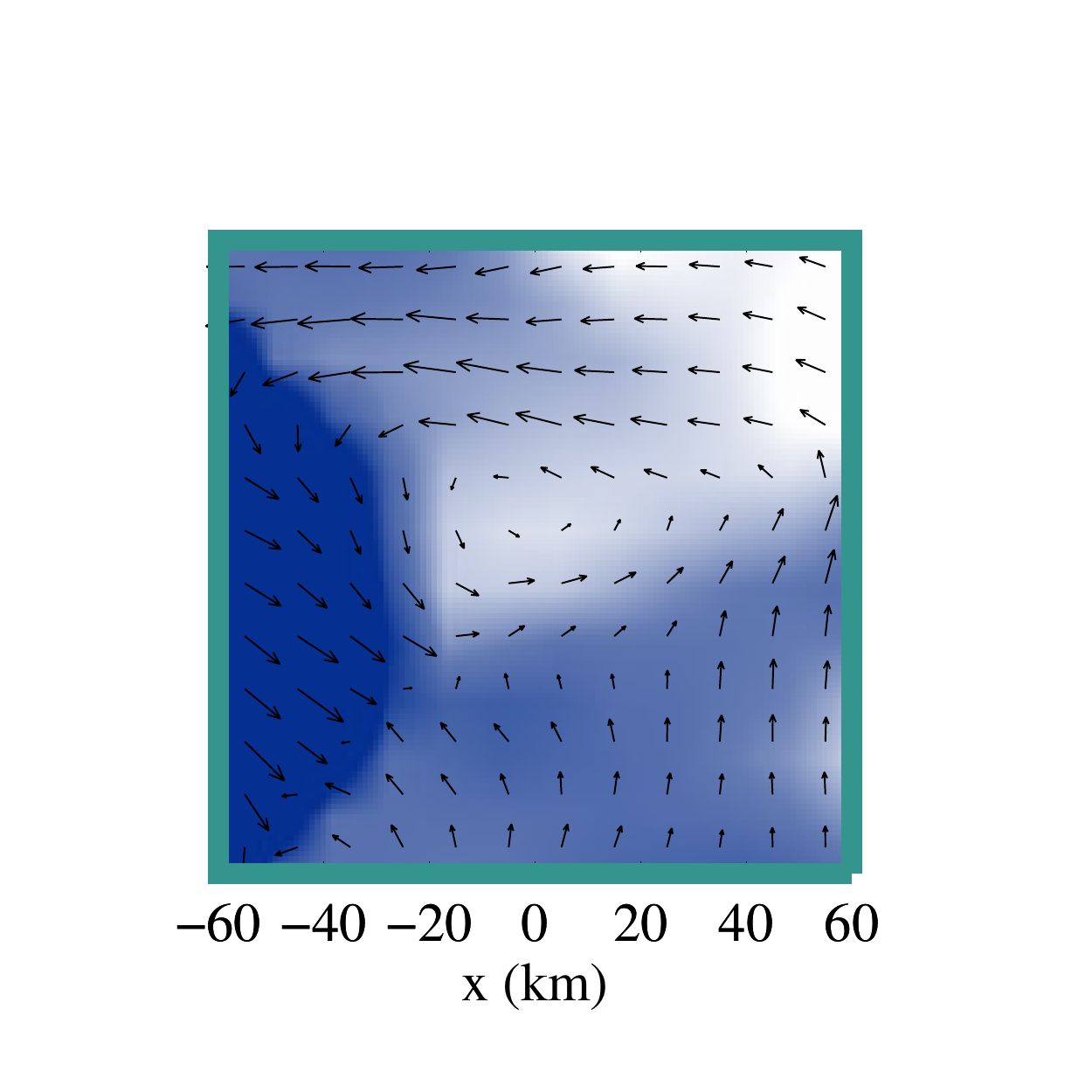}
\includegraphics[scale=0.32,trim=0 1cm 4cm 1.8cm]{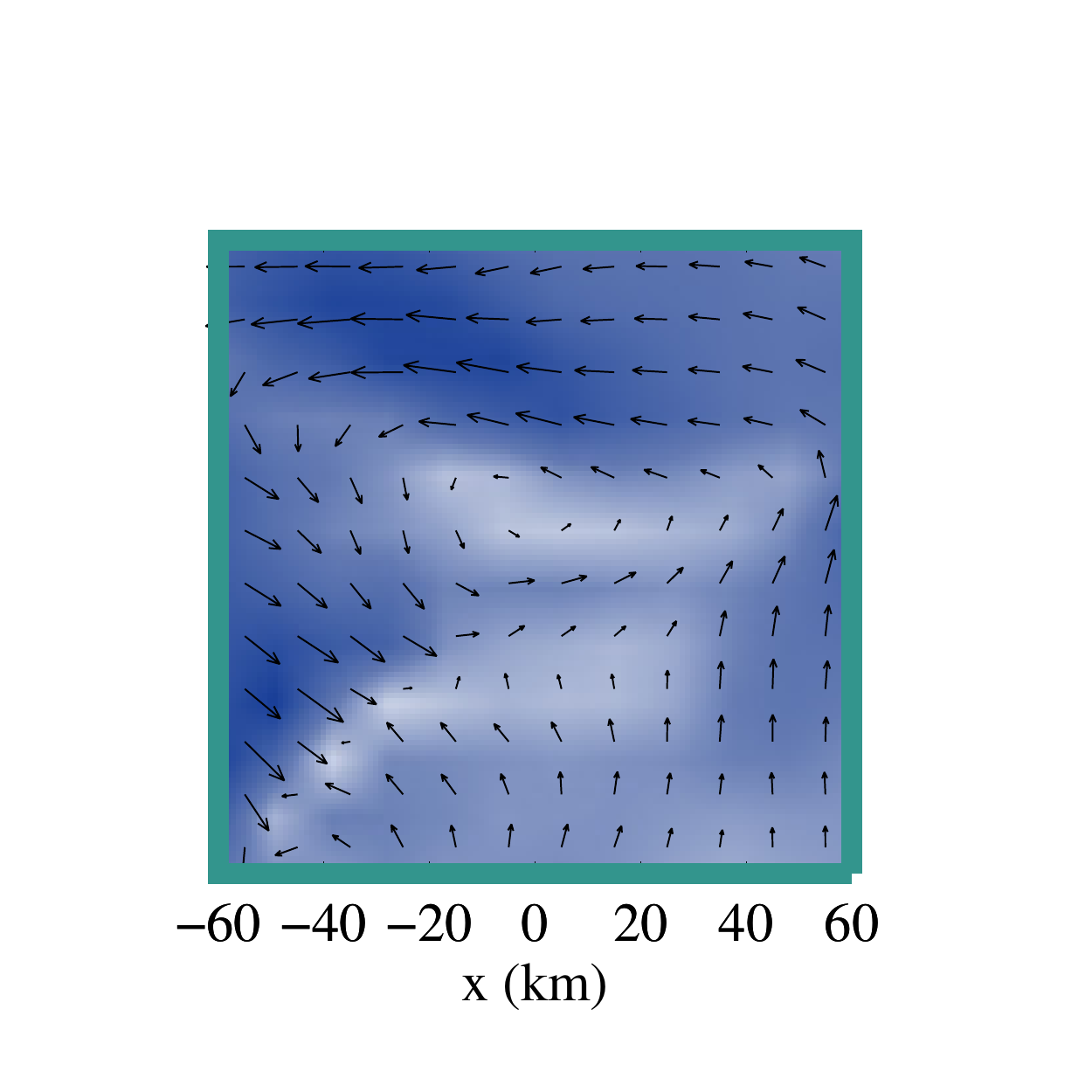}\\
\includegraphics[scale=0.32,trim=2cm 3cm 4cm -1.cm]{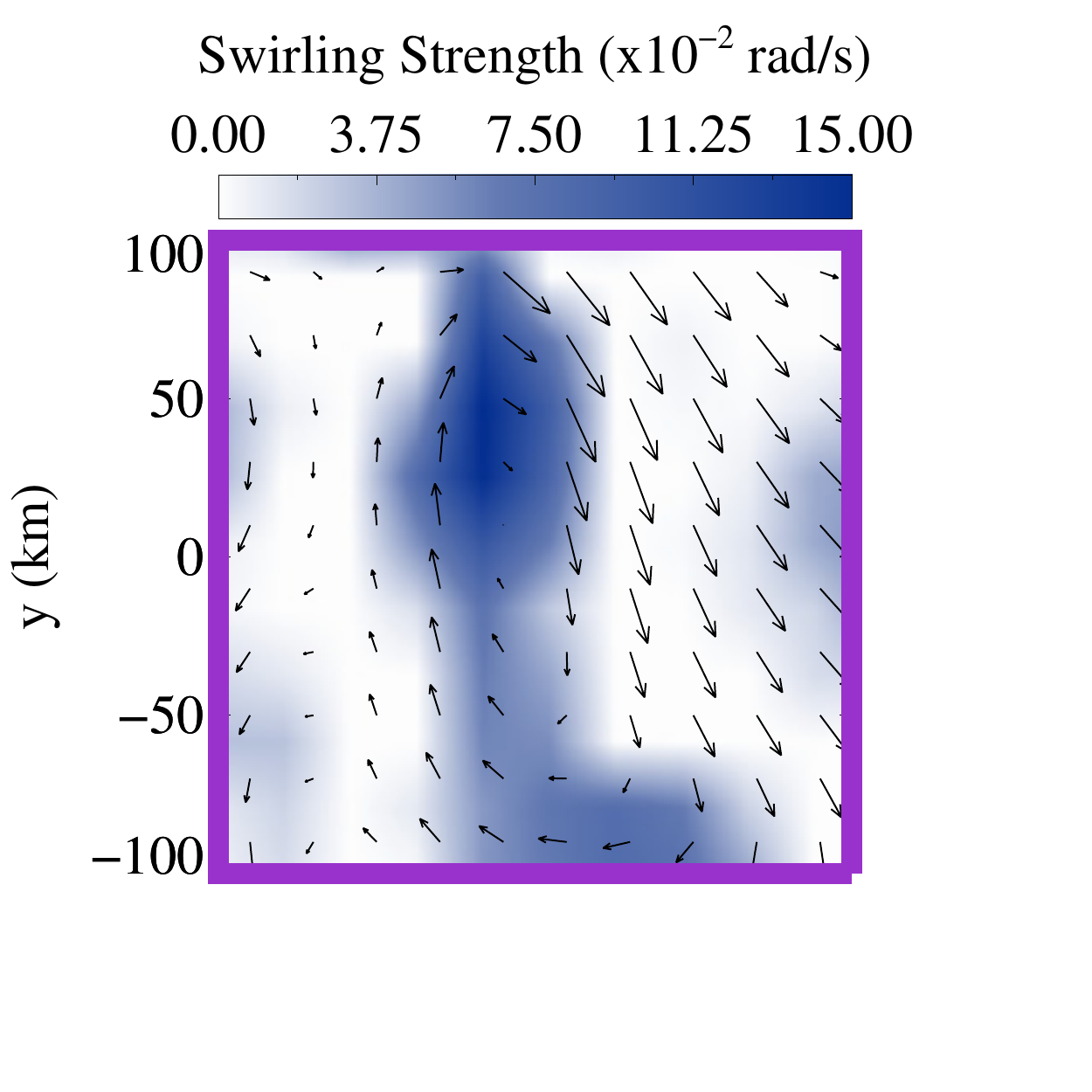}
\includegraphics[scale=0.32,trim=0 3cm 4cm -1.cm]{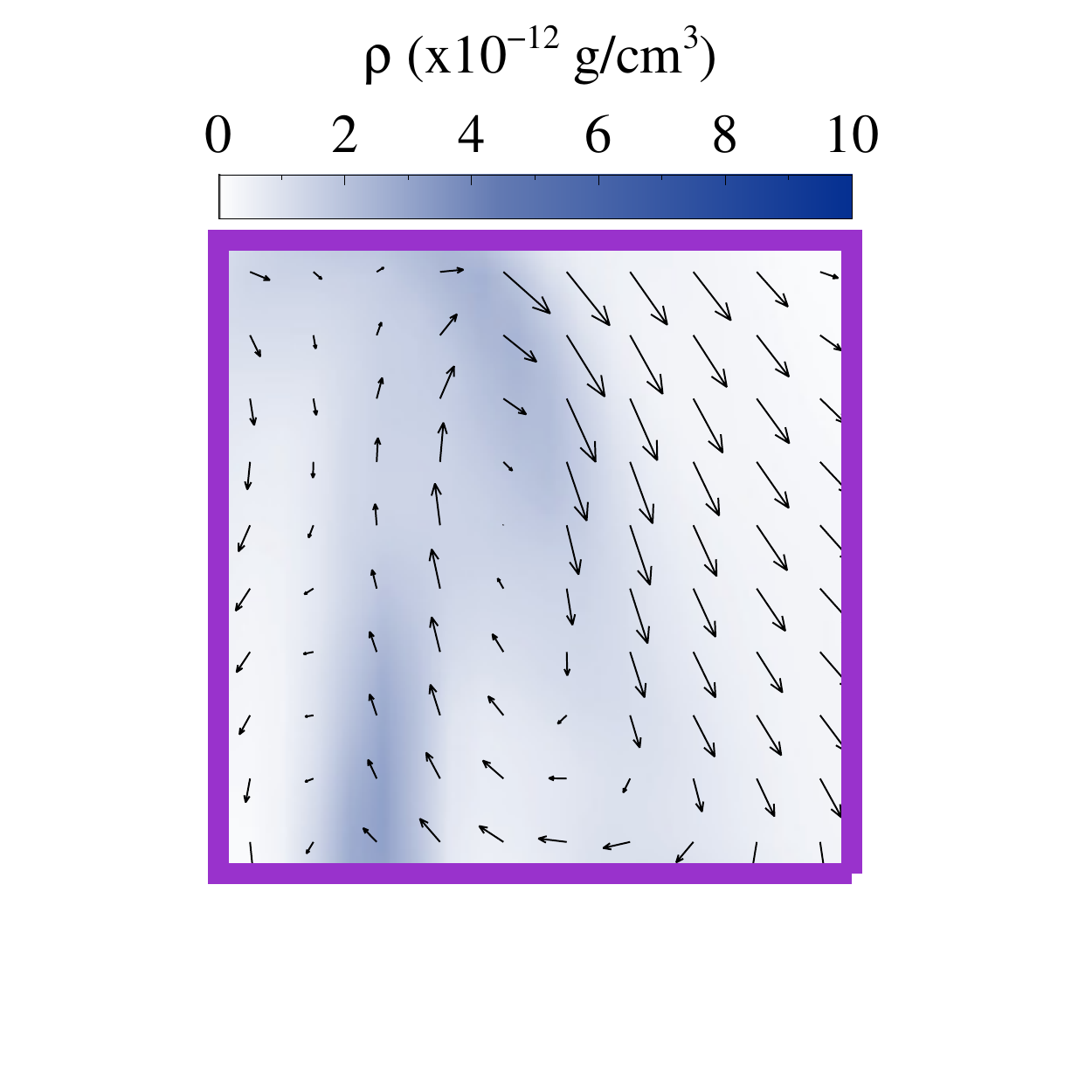}
\includegraphics[scale=0.32,trim=0 3cm 4cm -1.cm]{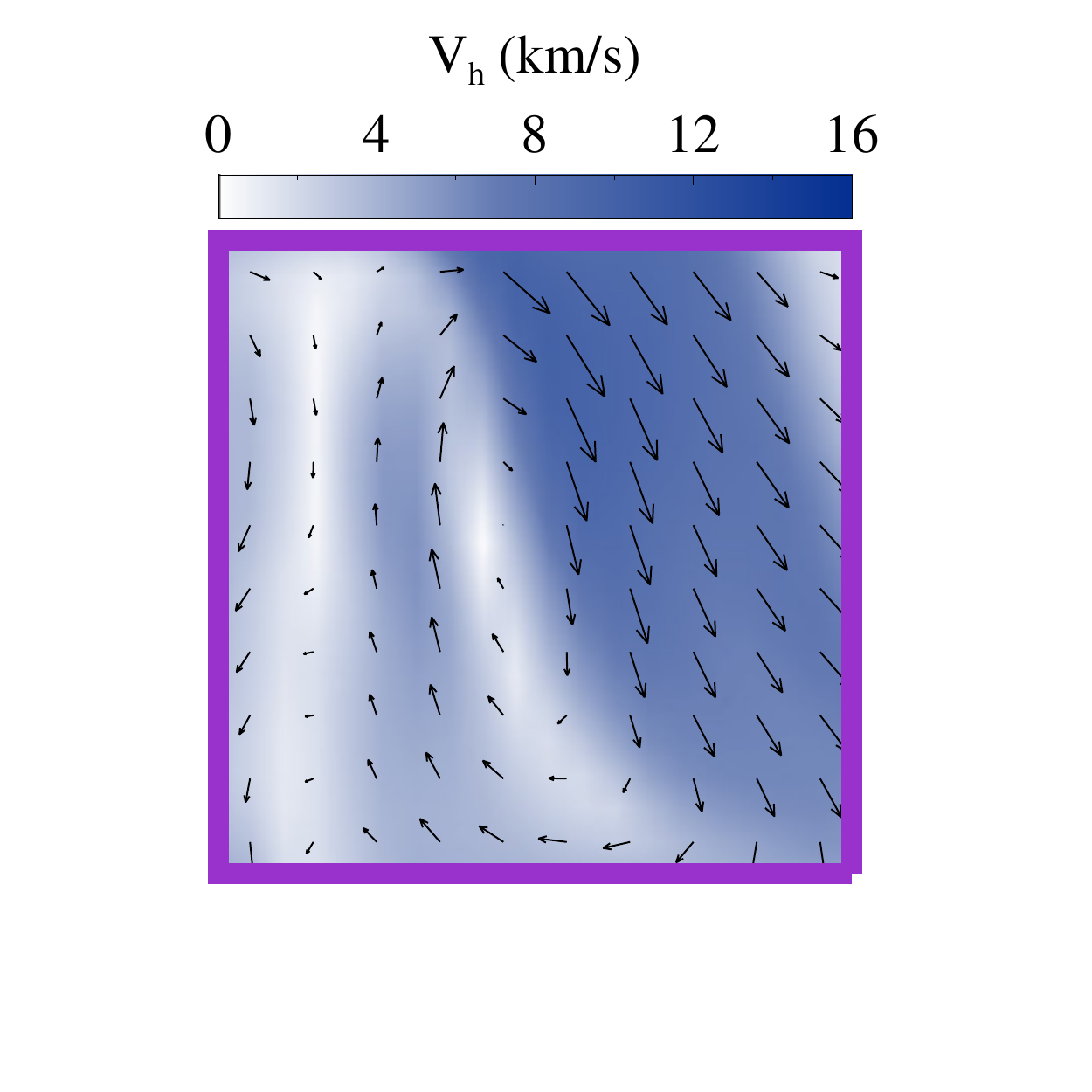}\\
\includegraphics[scale=0.32,trim=2cm 3cm 4cm 1.8cm]{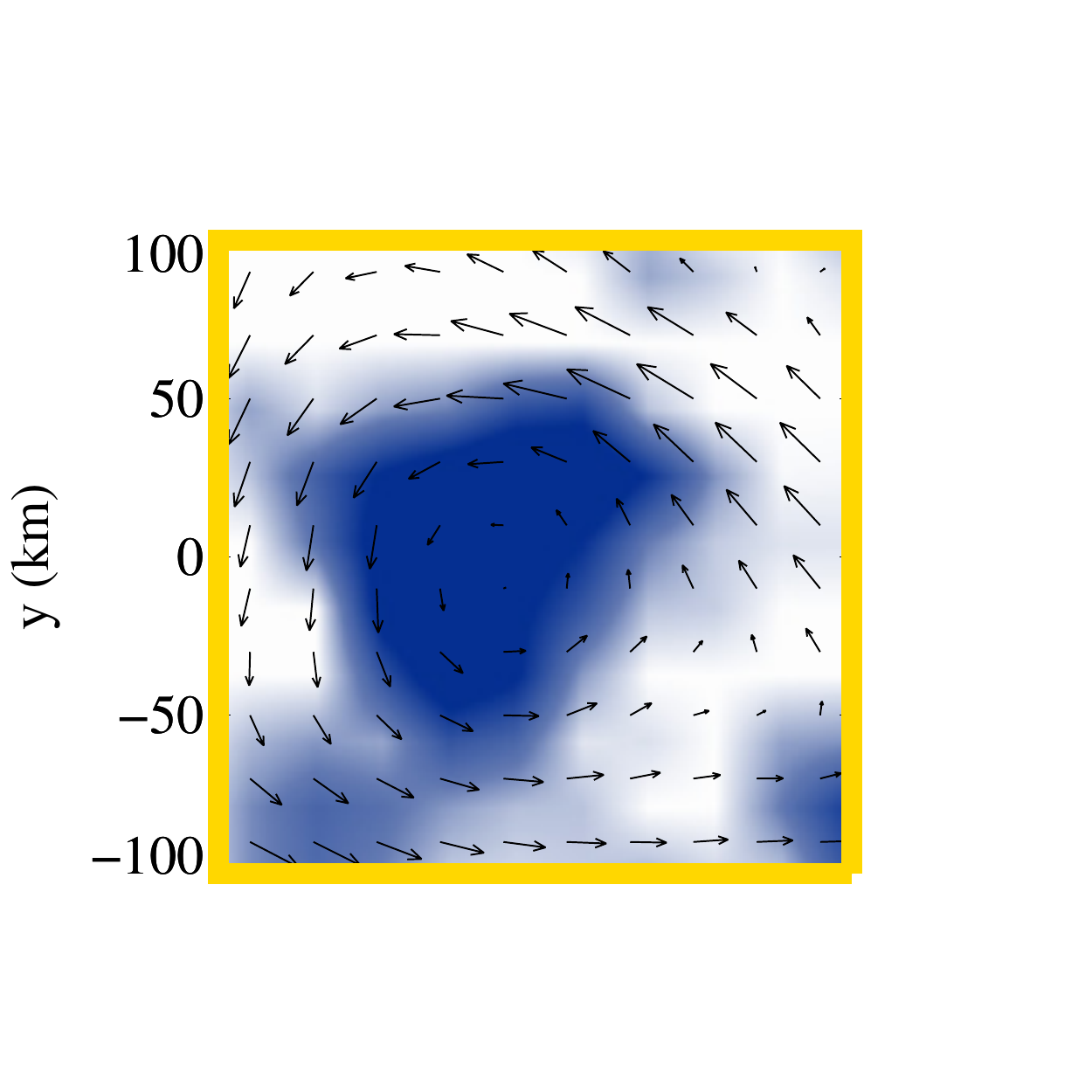}
\includegraphics[scale=0.32,trim=0 3cm 4cm 1.8cm]{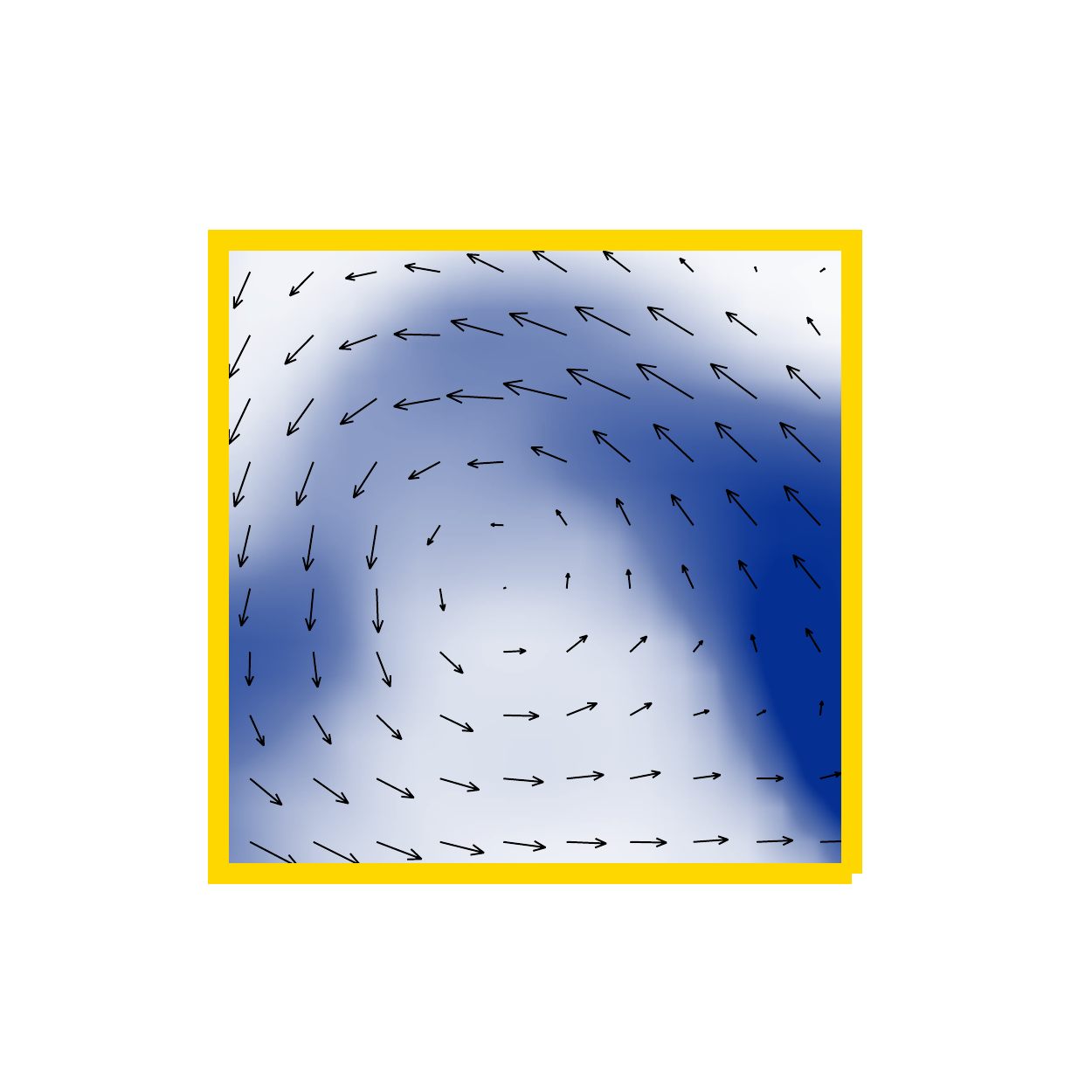}
\includegraphics[scale=0.32,trim=0 3cm 4cm 1.8cm]{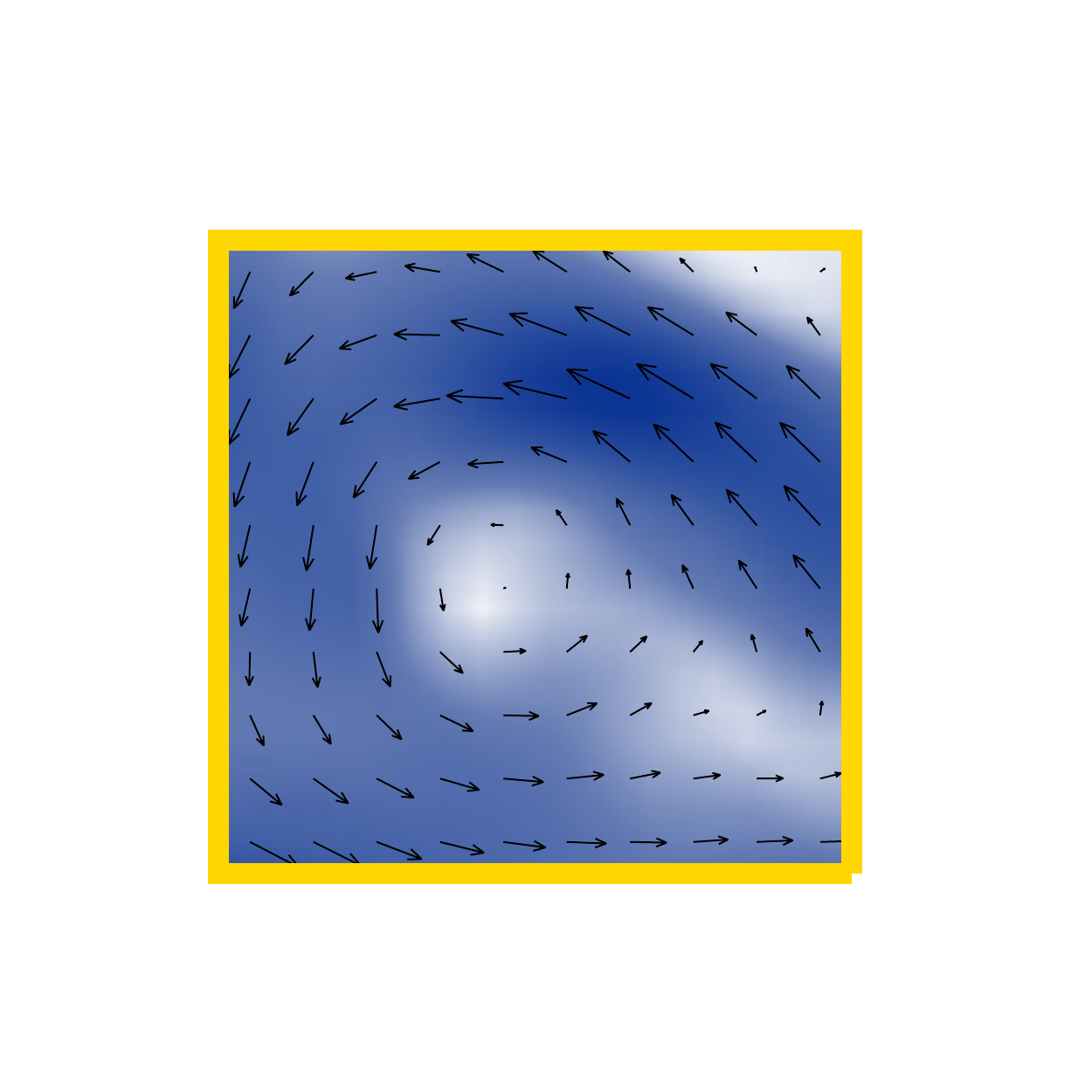}\\
\includegraphics[scale=0.32,trim=2cm 1cm 4cm 1.8cm]{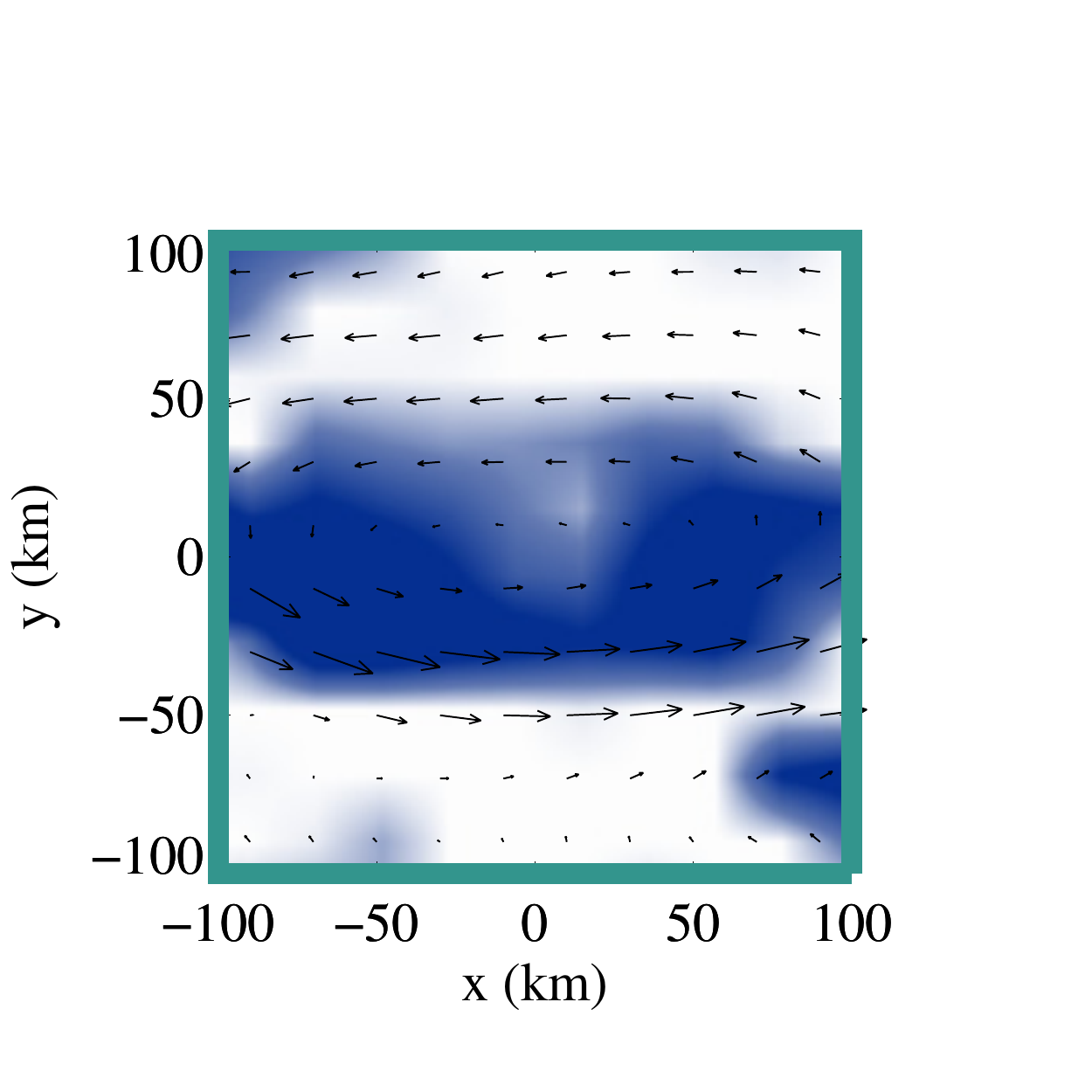}
\includegraphics[scale=0.32,trim=0 1cm 4cm 1.8cm]{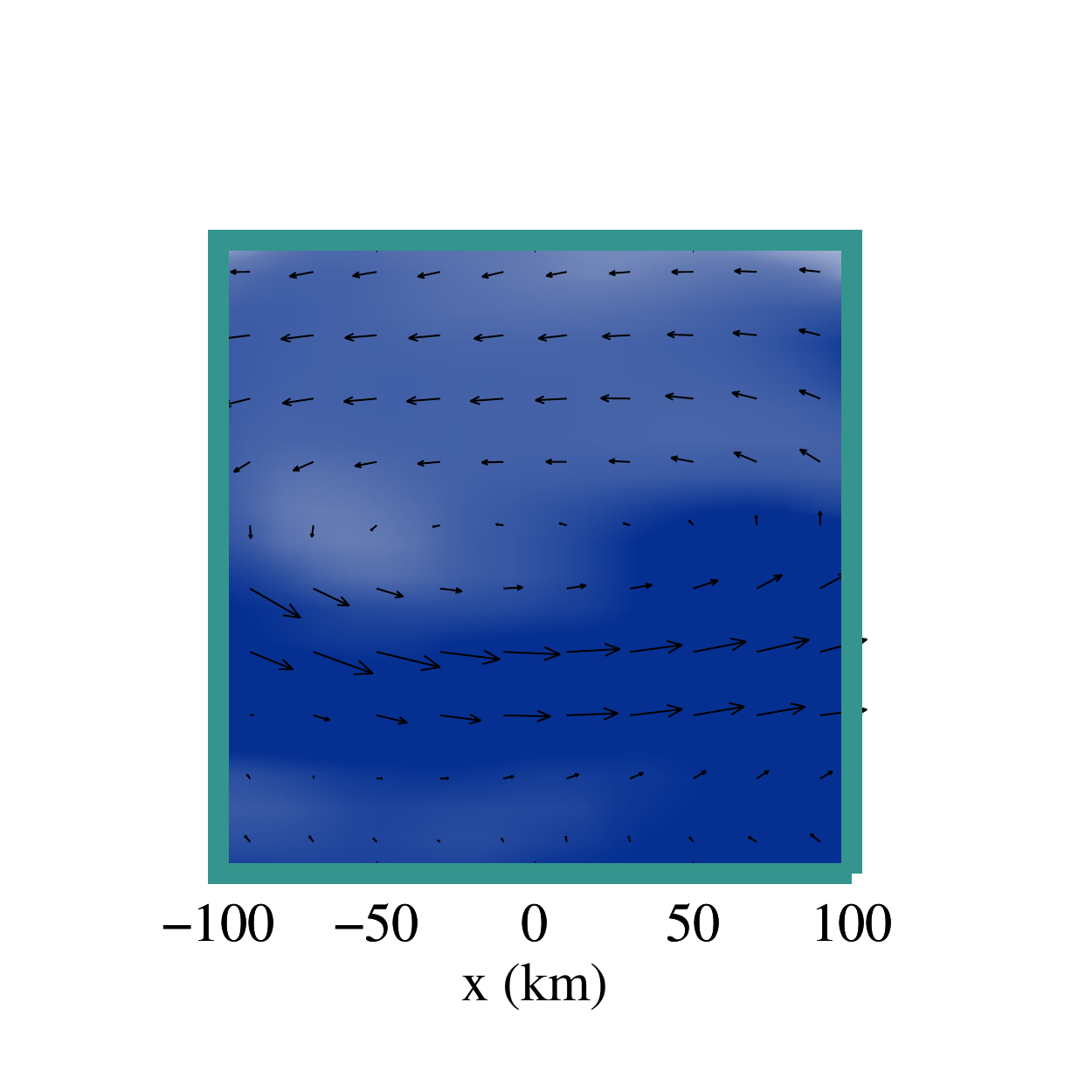}
\includegraphics[scale=0.32,trim=0 1cm 4cm 1.8cm]{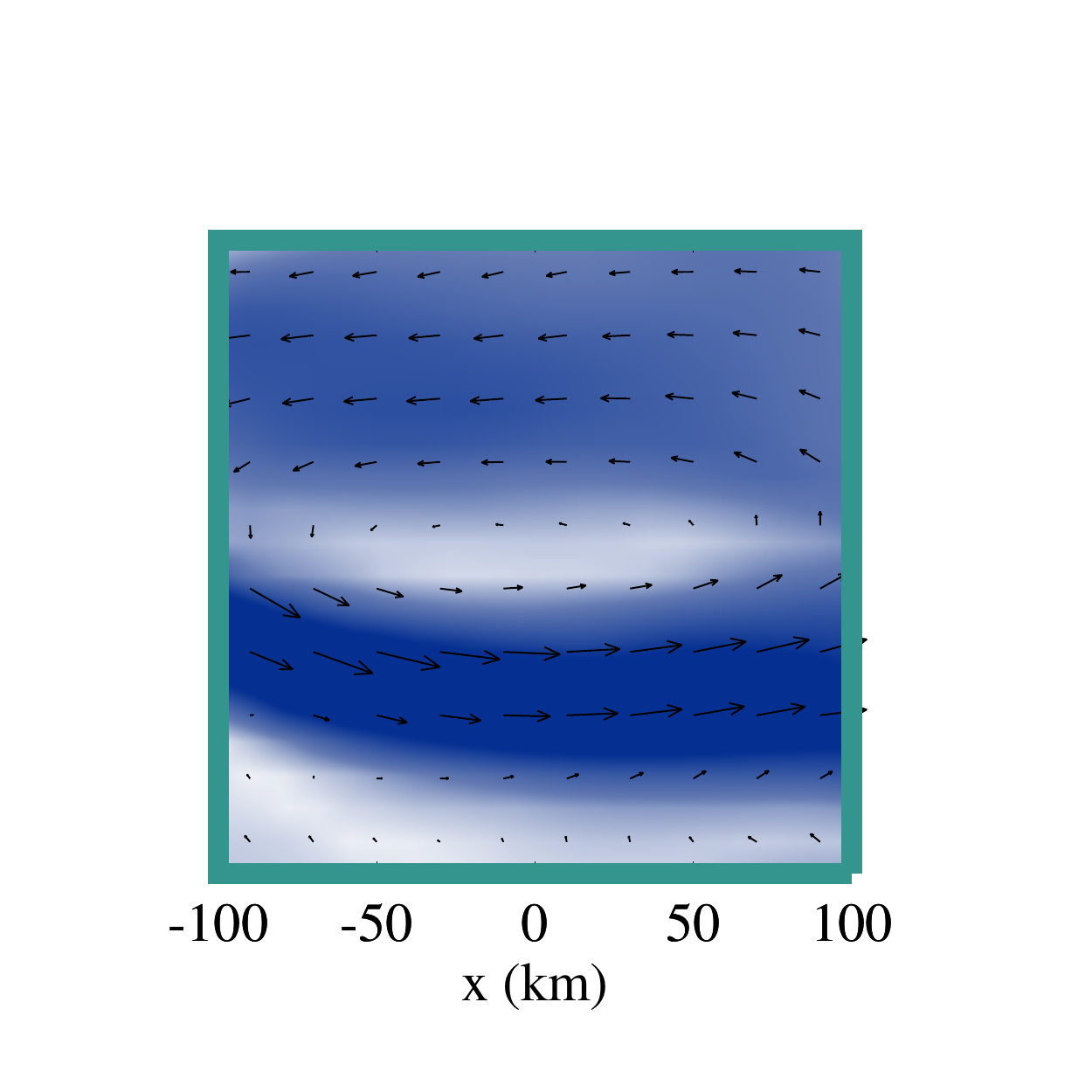}
\caption{Left to right: Profiles of swirling strength, mass density and horizontal velocity for three vortices (different rows with border colors corresponding to those of the squares outlining the vortices  in Fig. \ref{Fig10} on the solar surface (rows 1 to 3) and at 1.5 Mm above the surface (rows 4 to 6).
The over-plotted arrows represent the horizontal velocity. The longest arrow corresponds to 4 km/s and 16 km/s in the photosphere and the chromosphere, respectively.}
\label{Fig11}
\end{figure}

Figure \ref{Fig8} displays the vertical component of flow velocity at the $\tau=1$ layer (same as in Fig. \ref{Fig3}) with over-plotted contours of the vertical component of magnetic field strength ranging from 400 G to 2500 G. 
To investigate the characteristics of the individual vortices in detail, an area with a horizontal extent of $1.4\; \mathrm{Mm}\times1.4\;\mathrm{Mm}$ encompassing a magnetic element is selected (marked by a red square in the figure).
The three-dimensional sub-domain of the simulation box corresponding to the red square is displayed on the right.
This sub-domain extends 2.5 Mm in the vertical direction above the mean solar surface.
The map of the vertical component of magnetic field strength at the mean solar surface is also displayed (saturated at $\pm$ 400 Gauss).
Though there are numerous vortices present in this selected region, we have selected three vortices for visualization using VAPOR (\citealt{clyne2005}; \citealt{clyne2007}).
Velocity streamlines over these vortices are displayed in sea-green color, indicating fluid rotation.
In contrast, the magnetic field lines, displayed in deep purple, are aligned with the axes of the vortex cores.
The various physical properties of these three vortices in this magnetic element are examined and shown in Figs. \ref{Fig10} and \ref{Fig11}.
Though we present results for a single magnetic element, several other magnetic elements are examined and are found to show qualitatively similar behavior.

Fig. \ref{Fig10} displays the spatial profiles of various physical quantities in the selected magnetic element at the mean solar surface (left column) and  at 1.5 Mm above the surface (right column).
Vortices are outlined in black.
Note that, we apply a swirling strength threshold to detect vortices and swirling strength increases with height (will be discussed in the description of Fig. \ref{Fig11}).
Therefore, we miss the comparatively smaller and slower rotating photospheric counterparts for many chromospheric vortices.
However, in the photosphere, there are many vortices detected in the surroundings of the magnetic element where mixed-polarity magnetic fields exist.
As these mixed-polarity regions are highly turbulent and are strongly influenced by external flows, they undergo continuous flux cancellation and reconnection.
Therefore, vortices originating in these regions are extremely short-lived (a few seconds) and do not extend higher up in the atmosphere.
In contrast, the vortices that originate inside the magnetic concentration extend up to the chromosphere following the magnetic field lines and may transport energy into the chromosphere (\citealt{yadav2020}).
Three squares are shown in each panel at the locations of the selected vortices.
The areas of these squares (i.e. cross-sectional areas of the vortices) grow from the photosphere (left column) to the chromosphere (right column) due to the expansion of the magnetic flux tube hosting the vortices with height.
Also, the locations of the vortices are not exactly the same at the two heights because the cores of these vortices are curved and obliquely directed. 

Row (a) of the figure displays the vertical component of the magnetic vector.
The magnetic field is concentrated in the photosphere, whereas in the chromosphere it is almost homogeneous (the field varies by less than 20 \%\ in the plotted region). 
Mass density maps are displayed in row (b).
Vortices originating in the strong magnetic field concentrations have a lower density in the near surface layers because strong magnetic regions are evacuated due to pressure balance. 
In the chromosphere, however, they capture the neighboring plasma and we see high-density intrusions, particularly in the lower half of the plotted panel (see, e.g., the green square).
In row (c) of the figure, maps of the vertical component of the current are displayed.
In the chromosphere, currents have a tendency to be stronger where there are more vortices and they are primarily generated at the boundaries of the vortices due to shearing of the magnetic fields.
Also, the regions which are void of vortices have negligible currents.
Temperature maps are displayed in row (d). In the photosphere, vortices have lower temperatures at a given geometric height 
\footnote{at equal optical depth, such locations are often hotter, however  (e.g., \citet{sami1993})} as they are in the strong magnetic concentration, while in the chromosphere, they often have high temperatures and heating appears to be local and co-located with current sheets in row (c).

To analyze the vortex characteristics in detail, the three vortices shown in Fig. \ref{Fig8} are investigated individually.
The maps of various physical quantities over vortices with over-plotted velocity vectors are shown in Fig. \ref{Fig11}.
Here, the top three rows correspond to the mean solar surface while the bottom three rows correspond to the chromosphere. 
The spatial extents of vortices are different at these two layers as the vortices expand with height and have a larger cross-sectional area in the chromosphere than in the photosphere.
The color of the borders of different panels refers to the color of the squares in Fig. \ref{Fig10}.
The first column displays the swirling strength profiles.
Swirling strength increases radially towards the center of each vortex which implies that the angular velocity increases, indicating that the vortices do not rotate like a rigid body.
Also, since the average mass density is lower in the chromosphere than in the photosphere, vortices rotate faster and have relatively higher values of swirling strength (angular velocity) in the chromosphere in comparison to the photosphere.
In addition, their shape may evolve quite strongly from the photosphere to the chromosphere.
Whereas one of the three vortices remains roughly circular in the shape over the whole height range, the other two are quite elongated in the chromosphere.
The second column displays the maps of mass density. 
Vortices that reach up to the chromosphere are located in regions of lower mass density (inside strong magnetic concentrations) at the mean solar surface, they nonetheless capture and trap plasma from the surroundings.
A similar pattern is visible in the three bottom rows corresponding to the chromosphere.
The maps of the horizontal flow velocity are displayed in the third column of the figure. 
At the solar surface as well as in the chromosphere, plasma is rotating faster near the edges than at the center in terms of absolute speed. However, the angular velocity is larger closer to the center.
%%%%%%%%%%%%%%%%%%%%%%%%%%%%%%%%%%%%%%%%%%%%%%%%%%%%%%%%%%
     \begin{figure*}
        \centering
      \includegraphics[scale=0.45, trim =0 1.cm 0 0]{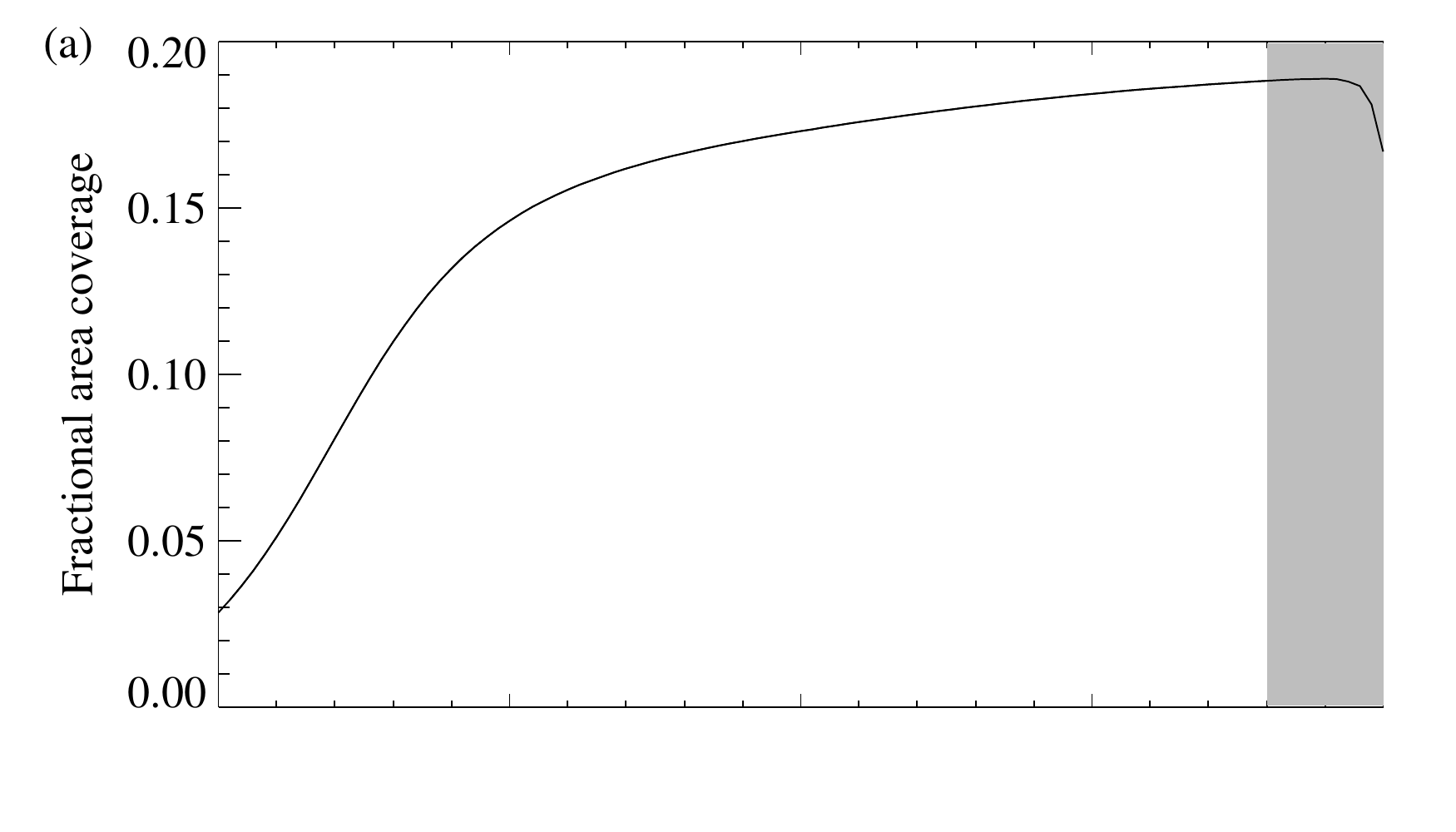}
     \includegraphics[scale=0.45, trim =0 1.cm 0 0]{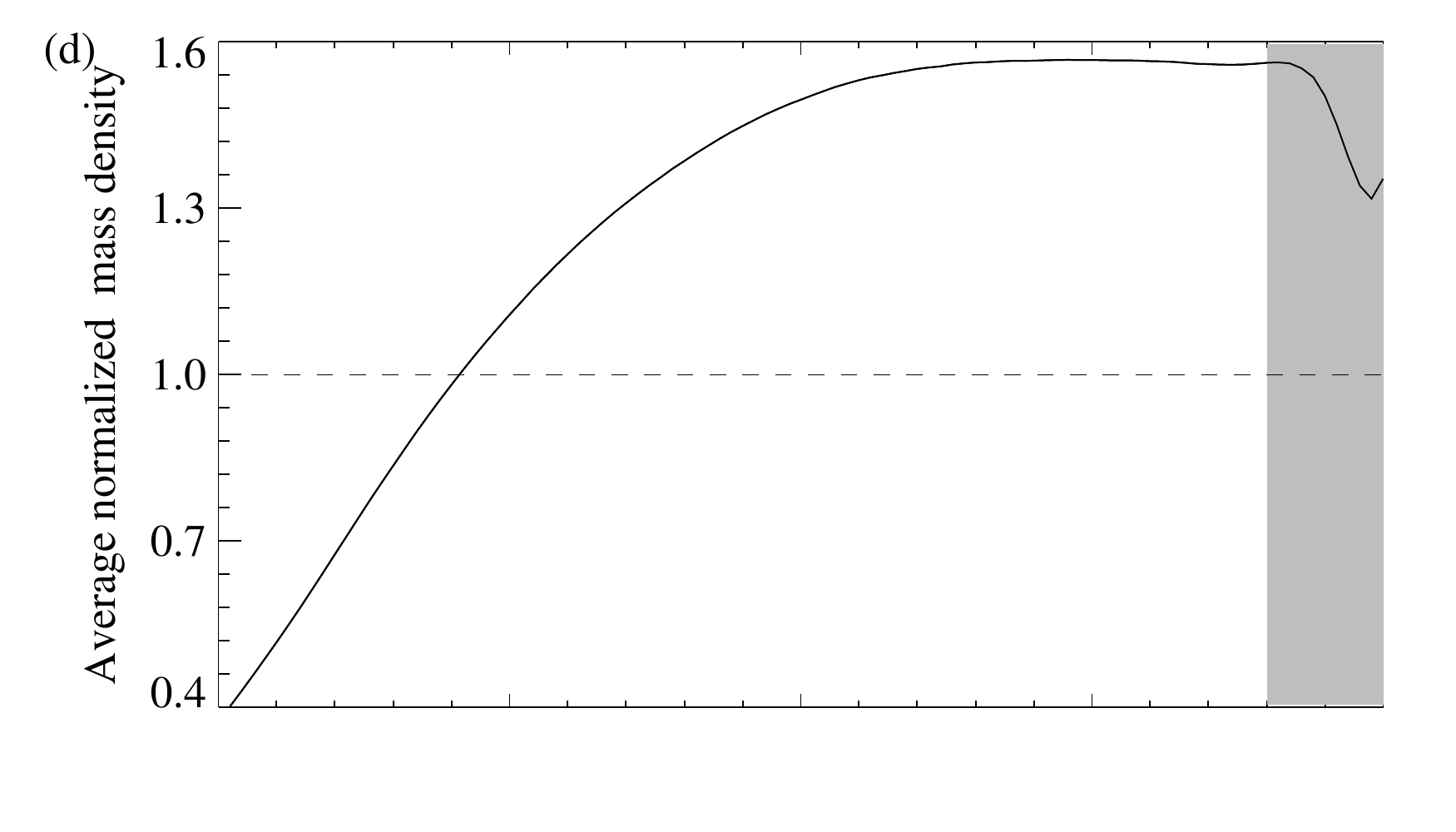}\\
 \includegraphics[scale=0.45, trim =0 1.cm 0 0]{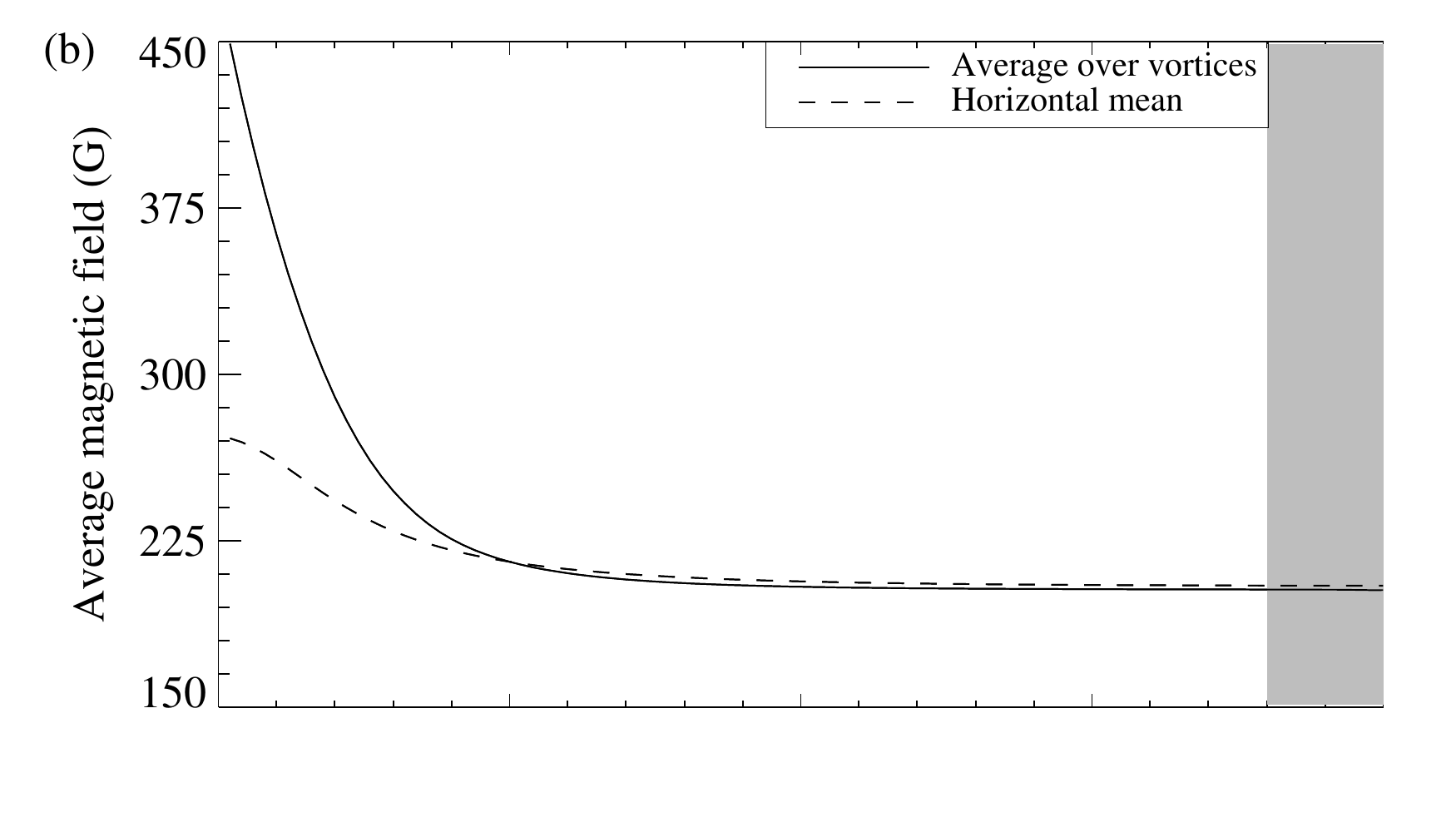}
   \includegraphics[scale=0.45, trim =0 1.cm 0 0]{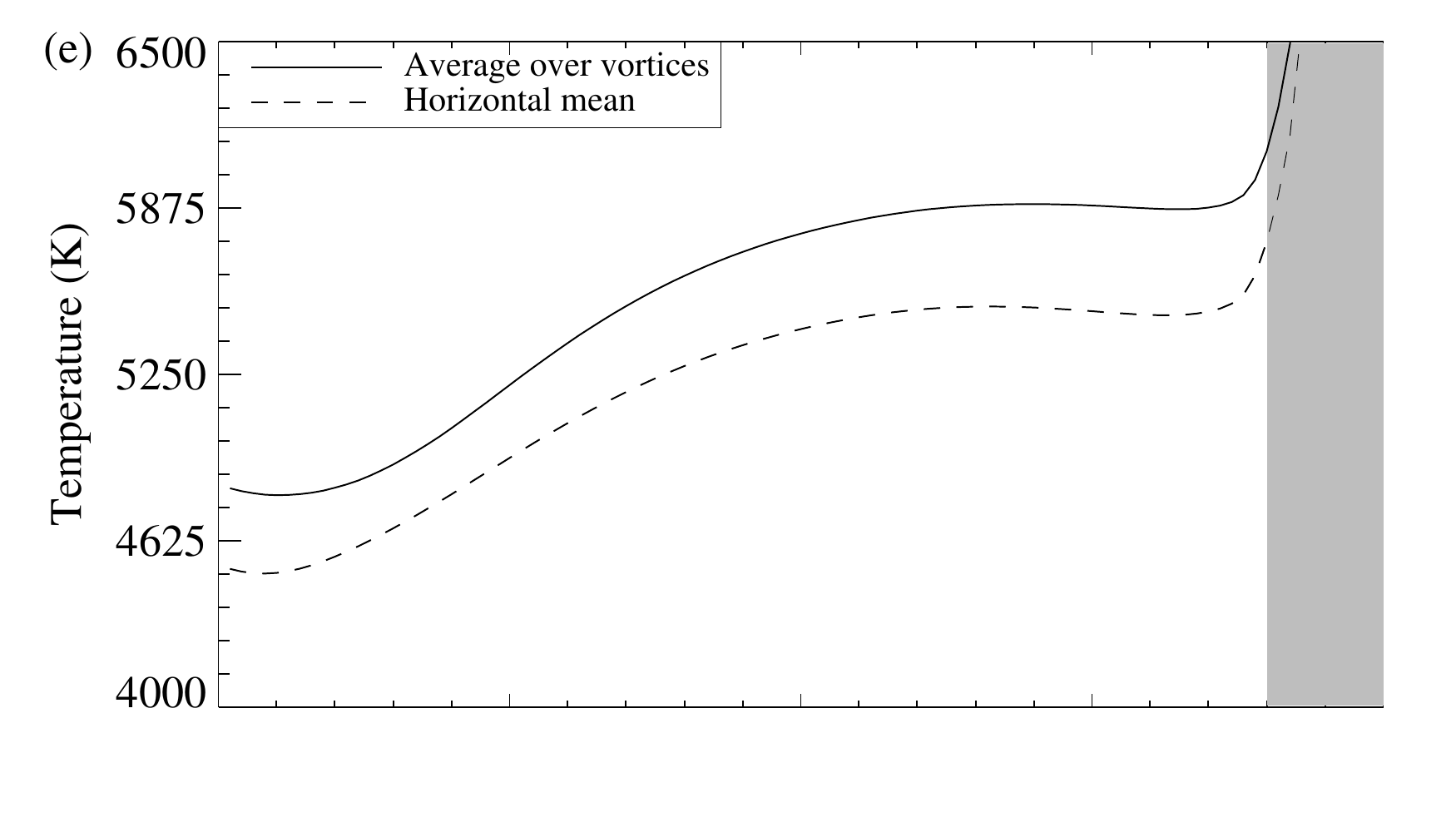}\\
  \includegraphics[scale=0.45, trim =0 0.2cm 0 0]{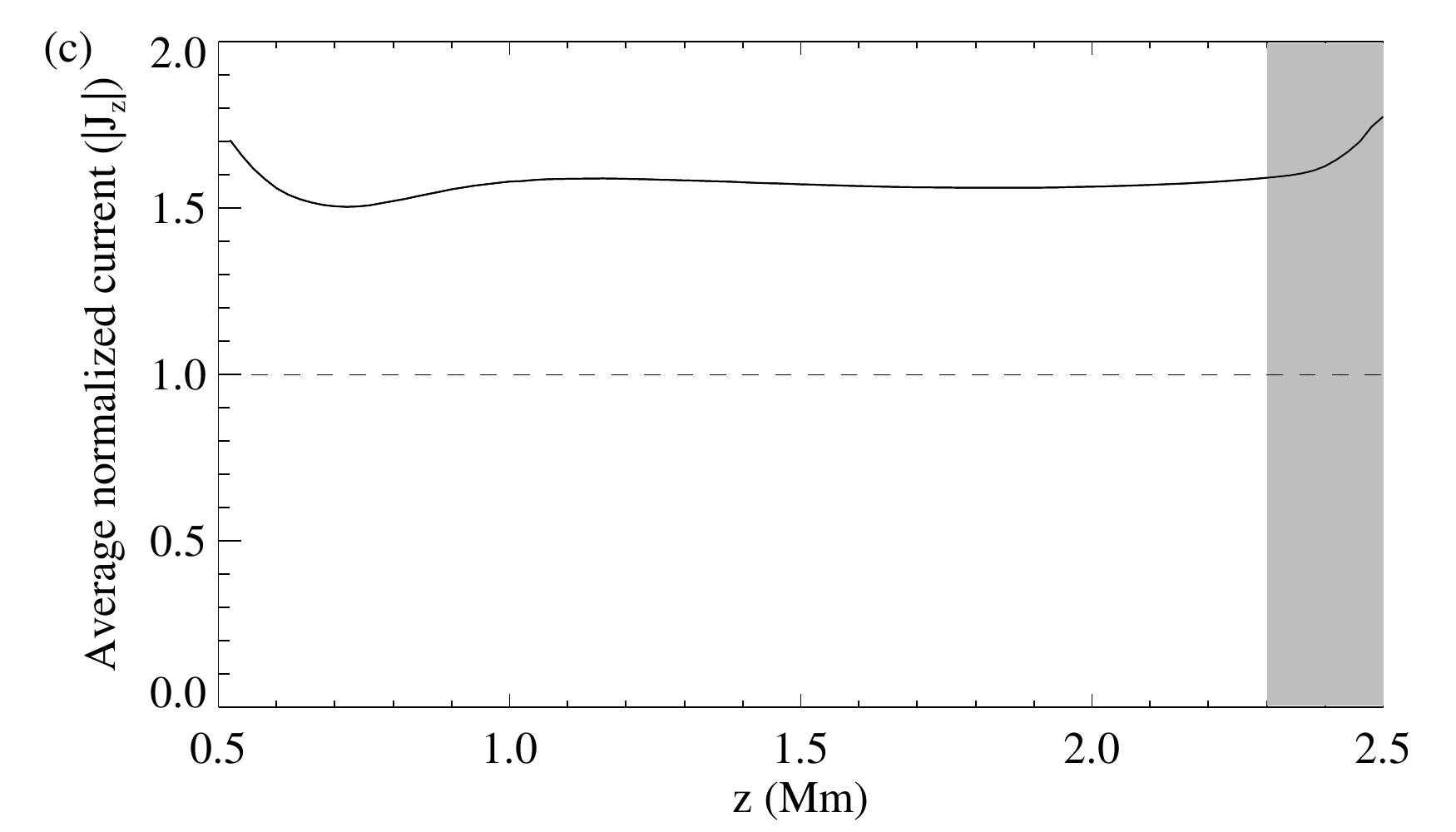}
   \includegraphics[scale=0.45, trim =0 0.2cm 0 0]{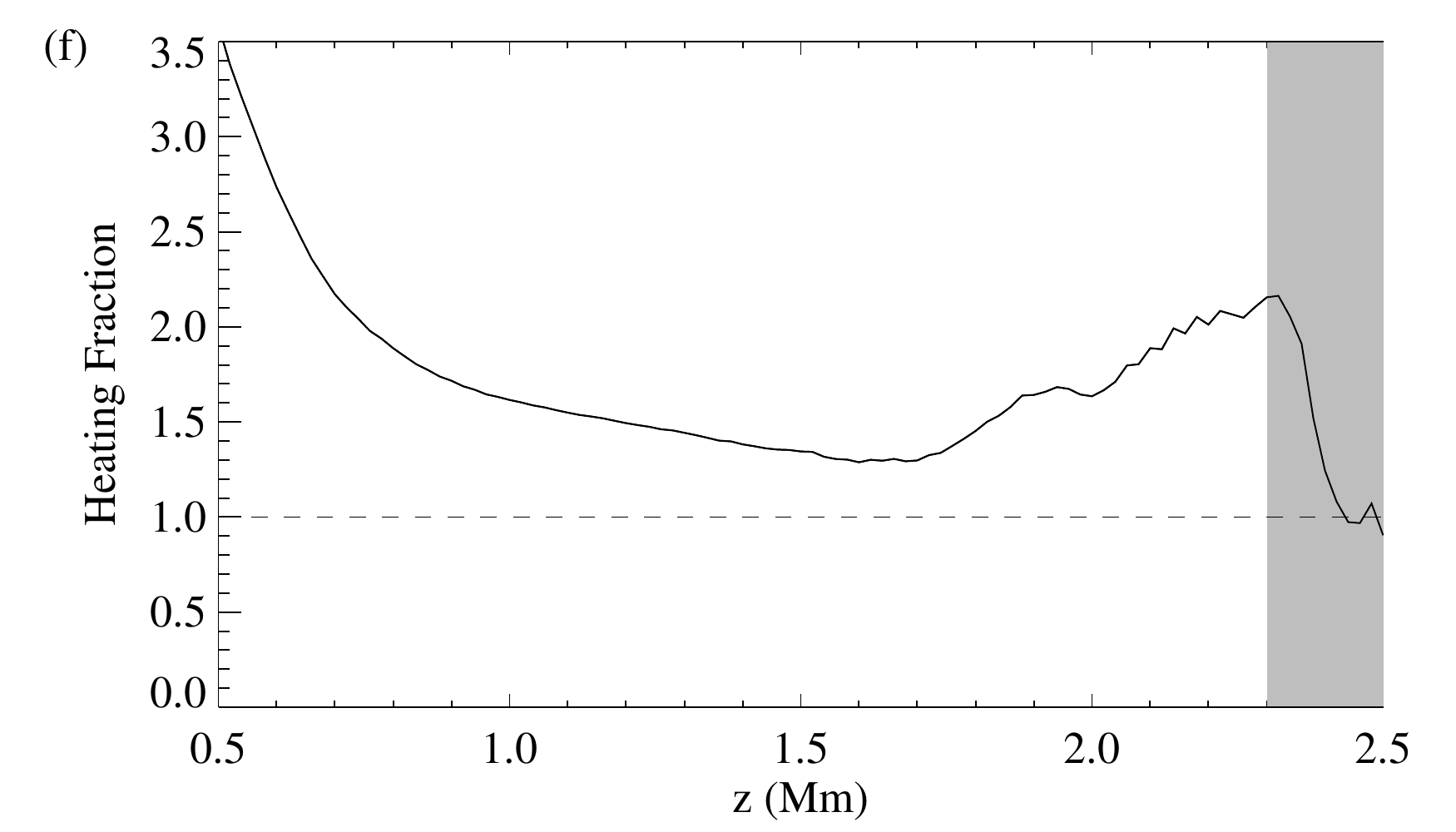}\\
   \caption{Dependence of vortex properties with height. (a): Area fraction (fill factor) covered by vortices, (b): Horizontally averaged magnetic field strength in vortices (solid) and over the whole domain (dashed), (c): Horizontally averaged  unsigned vertical component of current in vortices (normalized to the density averaged over the whole domain), (d): Horizontally averaged mass density in vortices (similarly normalized), (e): Horizontally averaged temperature in vortices (solid curve) and over the entire domain (dashed curve) and (f): Horizontally averaged total heating (viscous+resistive) in vortices (normalized to the total heating averaged over the whole domain). The region near the top boundary is shaded Gray as this part may be affected by the imposed top boundary conditions and should be ignored.}
   \label{Fig7}
\end{figure*}
 %%%%%%%%%%%%%%%%%%%%%%%%%%%%%%%%%%%%%%%%%%%%%%%
  \begin{figure}
   \centering
          \includegraphics[scale=0.55,trim=1.cm 0 1.cm 0]{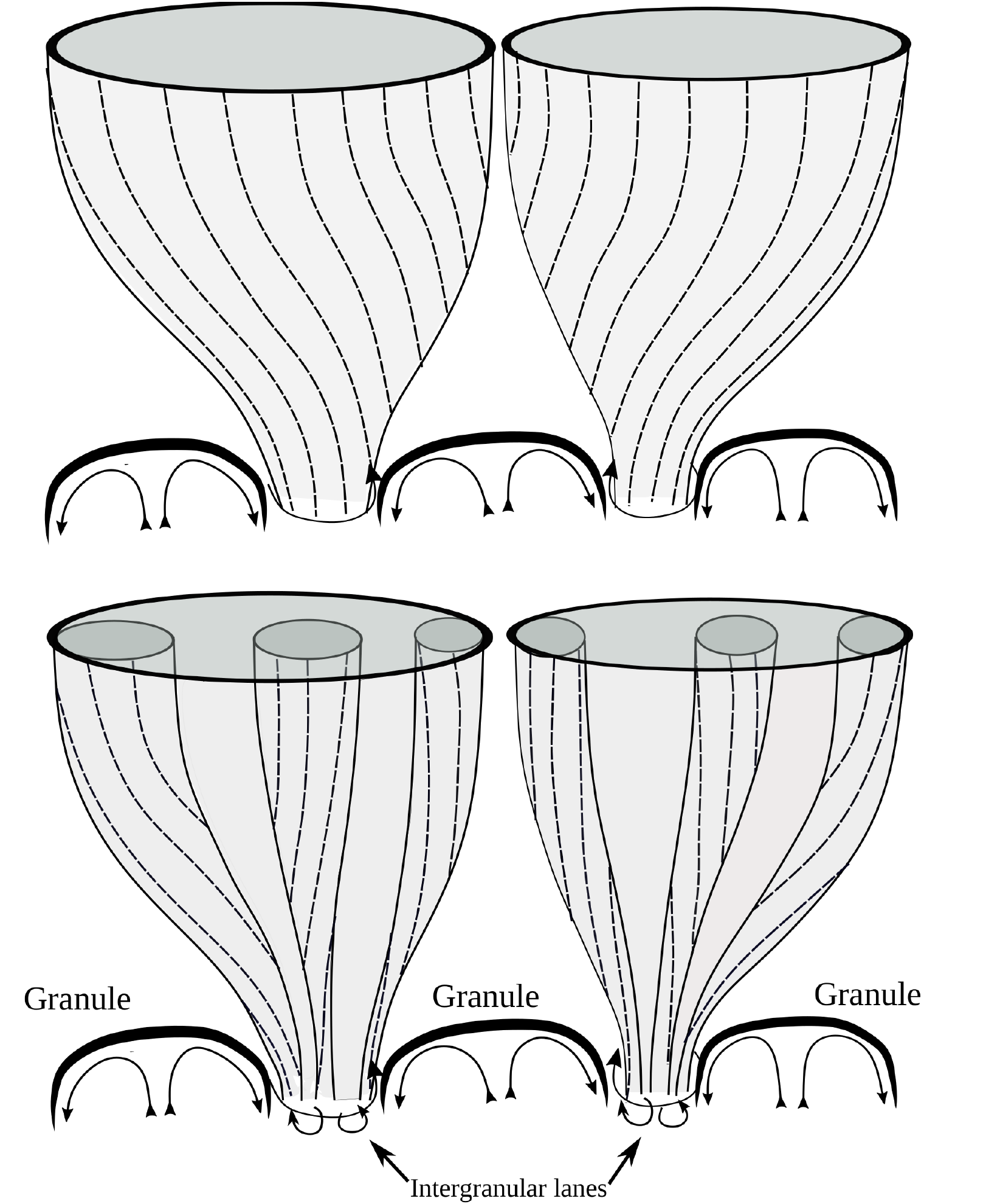}
      \caption{
      A simplified cartoon showing how turbulent motions drive vortices at different spatial scales. The lines with arrow heads represent the surface flows, being convective motions in the granules and small-scale turbulent motions in the intergranular lanes. These turbulent motions cover a range of scales and couple to the magnetic field (dashed curves)  producing vortices at various scales from the whole flux tube (upper panel) down to the resolution limit of the simulation (lower panel).}
         \label{Fig5}
    \end{figure}
 
\subsection{Statistical properties of vortices}
In a recent study, \citet{yadav2020} have demonstrated that vortices exist over a range of spatial scales and smaller vortices are the dominant carriers of energy. 
Therefore, in the present paper we concentrate on investigating the properties of small-scale vortices.
To study their general characteristics, a statistical investigation is performed by taking the temporal average for  42 snapshots covering a 7-minute long time sequence at 10 s cadence.
The vortices that we are investigating are small-scale (spatial size of $\sim$50-100 km near the solar surface and $\sim$100-200 km in the chromosphere) and short-lived (lifetime $\sim$20-120 s).
They are abundant throughout the simulation domain, therefore the time sequence of 7 minutes is sufficient for their statistical analysis. 
We display the dependence of vortex properties with height in Fig. \ref{Fig7}.
Note that the height scale starts at 0.5 Mm above the average solar surface, i.e. roughly at the base of the chromosphere.
Panel (a) displays the filling factor of vortices i.e. fractional area covered by vortices as a function of height.
In the lower chromospheric layers, area coverage increases with height because of the magnetic field expansion, however, in the upper layers magnetic field is nearly homogeneous (shown in panel (b)) and hence the filling factor of vortices does not increase much.
Panel (b) displays the height variation of the average magnetic field strength over vortices (solid) and horizontal mean (dashed).
Up to a height of $\sim$1 Mm, vortices have higher averaged magnetic field strengths as they originate inside the magnetic concentrations.
However, in the higher layers, the field strength is nearly homogeneous and the average magnetic field over vortices is nearly same as the overall horizontal mean. 
Panel (c) displays the vertical component of current (unsigned) averaged over vortices relative to the horizontal mean at the same geometrical height confirming the association of small-scale vortices with strong currents in the chromosphere, which we reported in Sect. \ref{individual}.
The mass density averaged over the vortices at a given height relative to the density averaged over the entire horizontal domain at the same geometrical height is displayed in panel (d).
Up to a height of $\sim$1 Mm, vortices have a lower mass density than the average mass density at the same geometrical height.
This is due to their occurrence in strong magnetic regions that are evacuated.
Above $\sim$1 Mm vortices have $\sim$50$\%$ higher density than the average at that height as vortices trap the plasma in the chromosphere.
Panel (e) shows the comparison of the average temperature over vortices (solid curve) with the temperature averaged over the whole horizontal plane (dashed curve) at each geometrical height.
The average temperature over vortices is always higher than the horizontally averaged temperature at the same geometrical height indicating their importance in the heating of the  chromosphere. 
Note that in lower layers (around $z=0$) the temperature in vortices is lower than the average, as they are located in magnetic elements that are cooler at equal geometrical height. 
Panel (f) displays the total heating (viscous + resistive) over vortices relative to the horizontally averaged heating.
Excessive heating over the vortices indicates their energetic importance, especially in the upper chromosphere.
The results obtained in the uppermost layers (Gray shaded region in all panels) should be ignored as they are affected by the closed top boundary condition and therefore, can be of non-physical origin.

In view of the results obtained we suggest the following formation mechanism for the small-scale vortices and display it in a simplified cartoon in Fig. \ref{Fig5}. 
Vortices are formed at various scales starting from the sizes of the intergranular lanes, possibly due to the angular momentum conservation of downflowing plasma.
Because of the turbulent nature of the plasma, vorticity cascades from the injection scale to the smaller scales.
The smallest scales that can be resolved in our simulations correspond to a multiple of the grid resolution.
In the upper convection zone and near-surface layers, where the ratio of thermal gas pressure to magnetic pressure, i.e., plasma- $\beta$, is high, gas dynamics dominates over magnetic fields and magnetic field lines are frozen into the plasma.
Vortical flows in this layer create a twist in the magnetic field. 
This twist in the magnetic flux tubes then propagates to higher atmospheric layers in the form of a torsional Alfv\'en wave.
In the upper layers, plasma- $\beta$ is low and plasma dynamics is dominated by magnetic fields.
Here, the twisted magnetic flux tubes make the surrounding plasma co-rotate forming a chromospheric counterpart of the photospheric vortices.
Because of the inherent turbulence, larger vortex flows cascade down to the smallest sizes determined by the grid resolution and numerical diffusivity.
Using vortex detection methods based on velocity gradients, we select the smallest vortices present in the system.
Vortices originate inside the magnetic elements and are aligned along the magnetic field lines (as shown in Fig. \ref{Fig2} and Fig. \ref{Fig8}).
These vortices do not appear in funnel-like shapes as observed in quiet Sun vortex observations (\citealt{Wedemeyer-Bohm2012,ParkS2016,kostas2018}) because vortices follow the magnetic field lines and the magnetic flux tubes in active regions do not expand as much as in the quiet Sun regions at least above the temperature minimum layer, where the magnetic elements merge (\citealt{pneuman_1986}).

 \section{Conclusions}\label{4}
Vorticies are ubiquitous in the solar atmosphere and are suggested to play an important role in the heating of the solar atmosphere (\citealt{Wedemeyer-Bohm2012,kostas2018,jiajia2018}).
However, even with the highest resolution observations available, the internal structure of the vortex tubes is not yet resolved.
Moreover, small-scale vortices are more abundant and are shown to contribute more energy flux in the solar atmosphere than comparatively larger vortex flows (\citealt{yadav2020}).
Thus it is of considerable relevance to investigate their physical properties using high-resolution numerical simulations.
With this aim, a three-dimensional radiation-MHD simulation of a plage region was carried out. 
Using swirling strength criterion vortices were detected and isolated. 
We report the occurrence of very small-scale vortices in a unipolar plage region.
With diameters of $\sim$50-100 km in the photosphere and $\sim$100-200 km in the chromosphere, these are the smallest vortices that can be resolved in our simulations. Due to a lack of spatial resolution, such small-scale vortices have not been detected in observations yet.

Vortices have a lower mass density than the mean density both in the photosphere and in the lower chromosphere, this is in agreement with the previous simulations that were limited to the photosphere and lower solar chromosphere (\citealt{Moll2011,Kitiashvili2013}).
However, in the higher atmospheric layers, they are denser structures than the surroundings.
This finding conforms with the fact that vortices are mostly observed as absorbing (dark) intensity structures when observed in chromospheric radiation. (\citealt{Wedemeyer-Bohm2012,ParkS2016,kostas2018}).

In chromospheric layers, the temperature averaged over vortices is higher than the temperature  averaged over the whole horizontal domain in the chromosphere (Fig. \ref{Fig7}).
Previously, \cite{moll2012} had also found an association of stronger vortices with increased temperatures (simulations limited up to 800 km above the mean solar surface).
Using simultaneous observations from SST/CRISP and IRIS, \cite{ParkS2016} have also shown small-scale quiet Sun vortices to have higher temperatures than the surroundings in the upper chromosphere.
A particularly important result of our present analysis is that we found thin and strong current sheets develop at their interface.
There seems to be a correspondence between large values of current densities and the vortex locations.
Hence we conjecture that heating at the vortex sites could be due to current dissipation in addition to viscous dissipation.
Nonetheless, the ratio of viscous and resistive heating will depend on the magnetic Prandtl number i.e. the ratio of kinematic viscosity to magnetic diffusivity(\citealt{axel_2014,Rempel2017,axel_2019}).

There are speculations that vortices can be the driving mechanism for dynamic jet-like features observed in the chromosphere (\citealt{Kuridze2016,Iijima2017}).
Though the present findings, viz. the correspondence of vortices with magnetic field lines, higher density and higher temperatures support their possible relationship with chromospheric jets, a more comprehensive study is needed to establish such correspondence.

\begin{acknowledgements}
The authors thank the anonymous referee for insightful comments on the manuscript. The authors acknowledge L. P. Chitta and D. Przybylski for useful discussions on various aspects of this paper. This project has received funding from the European Research Council (ERC) under the European Union’s Horizon 2020 research and innovation programme (grant agreement No. 695075) and has been supported by the BK21 plus program through the National Research Foundation (NRF) funded by the Ministry of Education of Korea. N. Y. thanks ISSI Bern for support for the team "The Nature and Physics of Vortex Flows in Solar Plasmas".
\end{acknowledgements}
 \bibliographystyle{aa} % stylefile aa.bst
  \bibliography{aa} % Yourfile.bib
% %%%%%%%%%%%%%%%%%%%%%%%%%%%%%%%%%%%%%%%%%%%%%%%%%%%%%%%%%%
 \end{document}